\documentclass[lettersize,journal]{IEEEtran}
\usepackage{multirow,flushend}
\usepackage{makecell, lipsum}
\usepackage{threeparttable}
\usepackage{colortbl}
\usepackage{booktabs}
\usepackage{pgfplotstable}
\usepackage{pgfplots}
\usepackage{amssymb} 

\usepackage{amsmath,amsfonts}
\usepackage{algorithmic}
\usepackage{algorithm}
\usepackage{array}
\usepackage[caption=false,font=normalsize,labelfont=sf,textfont=sf]{subfig}
\usepackage{textcomp}
\usepackage{stfloats}
\usepackage{url}
\usepackage{verbatim}
\usepackage{graphicx}
\usepackage{cite}
\definecolor{mygreen}{rgb}{0.7,1,0.7}
\definecolor{myblue}{rgb}{0.75,0.95,0.99}

\begin{document}

\title{Analysis and Validation of \\Image Search Engines in Histopathology}

\author{Isaiah Lahr$^{1,\dagger}$, Saghir Alfasly$^{1,\dagger}$~\IEEEmembership{Member,~IEEE} , Peyman Nejat$^1$, Jibran Khan$^1$, Luke Kottom$^1$, \\Vaishnavi Kumbhar$^1$, Areej Alsaafin$^1$, Abubakr Shafique$^1$, Sobhan Hemati$^1$, Ghazal Alabtah$^1$, \\Nneka Comfere$^2$, Dennis Murphree$^{1,2}$, Aaron Mangold$^3$, Saba Yasir$^4$, Chady Meroueh$^4$, Lisa Boardman$^5$, \\Vijay H. Shah$^{5,6}$, Joaquin J. Garcia$^4$, H.R. Tizhoosh$^{1,\ddagger}$~\IEEEmembership{Senior Member,~IEEE\\ \vspace{0.1in}
$^1$ Kimia Lab, Dept. of Artificial Intelligence \& Informatics, Mayo Clinic, Rochester, MN, USA\\
$^2$ Dept. of Dermatology, Mayo Clinic, Rochester, MN, USA\\
$^3$ Dept. of Dermatology, Mayo Clinic, Scottsdale, AZ, USA\\
$^4$ Dept. of Laboratory Medicine and Pathology, Mayo Clinic, Rochester, MN, USA\\
$^5$ Dept. of Gastroenterology, Mayo Clinic, Rochester, MN, USA\\
$^6$ Dept. of Physiology and Biomedical Eng., Mayo Clinic, Rochester, MN, USA\\
$\dagger$ equal contributions,  $\ddagger$ corresponding author
}
\thanks{Manuscript submitted February 3, 2024.}}



\maketitle

\begin{abstract}
Searching for similar images in archives of histology and histopathology images is a crucial task that may aid in \textcolor{black}{patient tissue comparison} for various purposes, ranging from triaging and diagnosis to prognosis and prediction. Whole slide images (WSIs) are highly detailed digital representations of tissue specimens mounted on glass slides. Matching WSI to WSI can serve as the critical method for \textcolor{black}{patient tissue comparison}. In this paper, we report extensive analysis and validation of four search methods  bag of visual words (BoVW), Yottixel, SISH, RetCCL, and some of their potential variants. We analyze their algorithms and structures and assess their performance. For this evaluation, we utilized four internal datasets ($1269$ patients) and three public datasets ($1207$ patients), totaling more than $200,000$ patches from $38$ different classes/subtypes across five primary sites. 
Certain search engines, for example, BoVW, exhibit notable efficiency and speed but suffer from low accuracy. Conversely, search engines like Yottixel demonstrate efficiency and speed, providing moderately accurate results. Recent proposals, including SISH, display inefficiency and yield inconsistent outcomes, while alternatives like RetCCL prove inadequate in both accuracy and efficiency. Further research is imperative to address the dual aspects of accuracy and minimal storage requirements in histopathological image search.
\end{abstract}


\section{Introduction} 
\IEEEPARstart{W}{ith} the surge of development in digital pathology over recent years, there has been an increase in demand for image search solutions to allow pathologists to quickly compare tissue images and learn from previously diagnosed and treated cases. This process is expected to resemble peer review and consultation to acquire second opinions. However, there are several challenges and roadblocks in image search in digital pathology archives that make creating an efficient and semantically accurate image search engine extremely difficult. The first issue is the extremely large nature of whole slide images (WSIs), which are often gigapixel files. So far, the ideal search engines must offer a method to ``divide'' or break up these WSIs into smaller more manageable ``patches'' or tiles. That implies that an ideal search engine must use a form of unsupervised algorithm to divide each WSI into a small set of patches. That only a few patches have to be selected to represent the WSI is a computational and storage requirement that would directly impact the adoption of search technologies in hospitals. Subsequently, to \emph{index} the WSI, the selected small set of patches are then often run through a pre-trained, or fine-tuned deep network that extracts feature vectors, also called \emph{embeddings} or deep features, from each patch. This brings us to the second issue regardless of the quality of the deep features. The ideal search engine should also offer an encoding scheme that allows for efficient storage and fast comparison of the deep features. Efficiency is not solely determined by speed but also by the amount of memory required to index images. The latter aspect has often been disregarded in literature, favouring fast search algorithms. 

\textcolor{black}{Different image search methods have been proposed in recent years. Some of them are complete search pipelines, whereas the others are  rather just some components of a search engine.} In this paper, we report the extensive analysis and validation of four frameworks and some of their variations for search in histopathology archives. We examine the bag of visual words (BoVW) \cite{zhu2018multiple}, Yottixel \cite{yottixel}, Self-Supervised Image Search for Histology (SISH) \cite{sish}, and Retrieval with Clustering-guided Contrastive Learning (RetCCL) \cite{wang2023retccl}, in several configurations using four private and three public datasets.
We first provide a brief overview of content-based image retrieval (CBIR) literature to establish the historical perspective. We then motivate the selection of these four search frameworks and explain how they work. We then report a comprehensive set of results to illuminate the algorithmic structure and efficiency and accuracy performance of each approach. This is the first rigorous and comparative analysis and validation of the most relevant image search frameworks in histopathology. We expect that the results reported in this paper will make clear where we stand in image search as a community and will help pathologists and other stakeholders in making decisions about which ones, if any, are intrinsically more tailored to clinical workflow and worth testing for research purposes.

\section{Image Search in Histopathology}

The field of information retrieval, particularly in the medical context, is a well-studied field and has undergone significant advancements \cite{baeza1999modern, yang2008boosting, singhal2001modern,goodrum2000image,lee2006beyond,averbuch2004context,lopes2022health,sridhar2015content, chen2022deep}. In recent years, Content-Based Image Retrieval (CBIR) in general and image search techniques have emerged as potentially indispensable tools in the field of histology and histopathology in particular, both for traditional light microscopy and digital pathology \cite{schaer2019deep,yottixel,sish,zhu2018multiple,hegde2019similar,wang2023retccl,zaveri2020kimia,akakin2012content}. \textcolor{black}{CBIR searches and retrieves images from a database based on their visual features. In contrast to using textual metadata like captions, keywords, or tags, CBIR systems analyze the visual content of images (i.e., the pixels) to identify objects’ similarities and differences. Generally, various features are extracted from images (e.g., color, texture, shapes) to index and organize the images in the database \cite{huang1998content, gudivada1995content,rui1998relevance}. In this work, we mainly use ``image search'' as equivalent to CBIR. } 

\textcolor{black}{CBIR applications for pathology have been investigated even before the emergence of digital pathology and deep learning \cite{furht1995content,wetzel1999evaluation, comaniciu1999image}. In those techniques, generally handcrafted features would be used to characterize pathology images \cite{shyu1998local, shyu1999testing}. Computer vision research provided a wealth of handcrafted features, such as SIFT \cite{wangming2008application}, SURF \cite{velmurugan2011content}, and LBP \cite{camlica2015medical}. It is accepted empirically that deep features are superior to handcrafted features \cite{alhindi2018comparing, tizhoosh2018representing, ma2021image}.}

Well-designed and efficient image search technologies may enable pathologists to effectively access, analyze, and compare large volumes of tissue images, enhancing the accuracy of diagnoses and treatment planning \cite{tizhoosh2021searching,kalra2020pan,zhang2014towards}. A thorough validation of image search solutions in histopathology in recent years seems to be missing. Looking at the vast and rich literature on CBIR, one may be inclined to include many methods in a validation study. However, considering specific requirements of medical CBIR  decreases the number of candidate image search solutions drastically. Bridging from general medical image analysis to digital pathology leaves us with a handful of candidates. We established only one condition to make the final selection: the image search candidates in histopathology worth examination should introduce, or at least employ a distinct ``divide'' strategy to be feasible in clinical settings; dividing a WSI into patches and selecting a small set of those patches is the indispensable prerequisite for \emph{\textcolor{black}{patient tissue comparison}} (one may also represent the entire WSI with one vector \cite{hemati2021cnn,fashi2022self,bidgoli2022evolutionary}). Pertinent to the validation process, it is of secondary importance whether a search strategy has introduced its own divide approach or borrows the divide from literature. Based on all these considerations, we selected four image search frameworks, namely BoVW \cite{zhu2018multiple}, Yottixel \cite{yottixel}, SISH \cite{sish} and RetCCL \cite{wang2023retccl}. 

As recent examples, we did not include SMILY \cite{hegde2019similar},   HSDH (histopathology Siamese deep hashing) \cite{alizadeh2023novel} and High-Order Correlation-Guided SelfSupervised Hashing-Encoding Retrieval (HSHR) \cite{li2023high} in our validation. \textcolor{black}{Table \ref{tab:excluded} provides an overview of the challenges with these methods.  The main practical reason for not evaluating these three methods is that their code and/or trained models are not publicly available such that we cannot reliability reproduce their results.} As well, they lack a divide stage \textcolor{black}{and follow a brute force approach to WSI representation}. HSDH is mainly focused on a Siamese network to process single patches. SMILY and HSHR assume all WSI patches should be processed, a task that requires prohibitively extensive resources as most WSIs consist of several hundred if not several thousand patches \textcolor{black}{(see Section \ref{sec:expsetup} where the relationship of storage and AI democratization is reviewed)}.

\begin{table}
\centering
\textcolor{black}{
    \caption{Excluded Search Methods. \textcolor{black}{The `brute force' approach does not follow a \emph{divide \& conquer} scheme but processes all patches of a WSI.}}
    \label{tab:excluded}
    \begin{tabular}{c|ccc}
    \toprule
       Methods  & Patching & Code & Model(s) \\
       \toprule
SMILY \cite{hegde2019similar} & Brute force & not available & not available\\
HSDH  \cite{alizadeh2023novel}& Brute force & not available & not available\\
HSHR \cite{li2023high}        & Brute force & available & not available\\
         \bottomrule
    \end{tabular}
    }
\end{table}

\subsection{BoVW}

\begin{figure*}[htbp!]
    \centering
    \includegraphics[width=1\textwidth]{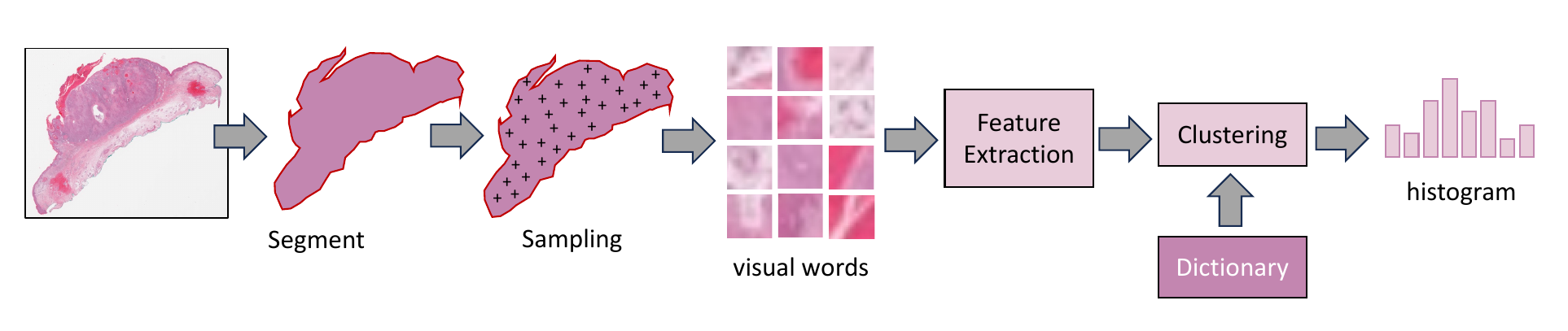}
    \caption{BoVW indexing pipeline.  Using a dictionary of visual words, the BoVW approach can represent a WSI with a histogram of visual words. Whereas various handcrafted features have been used in literature, herein, deep features are more likely to effectively represent visual words.}
    \label{fig:BoVW}
\end{figure*}

The Bag of Visual Words (BoVW) framework, often also called \emph{visual dictionary} approach, is generally quite well established and involves creating a dictionary by extracting and quantizing \emph{local descriptors} or features from small sub-images called \emph{visual words} \cite{csurka2004visual,sivic2005discovering,yang2007evaluating}. Different methods can be utilized for sampling and obtaining image descriptors. These methods may yield different image representations with varying discriminative abilities. The resulting image features are then used to construct a ``dictionary'' of visual words, which is employed for image encoding. BoVW generates visual word \emph{histograms} that incorporate both local and global information. The bag-of-features technique has been popular in computer vision applications, performing well in tasks such as image annotation, classification, and retrieval. It has also been applied to medical and biomedical image analysis, with notable examples achieving top performance in image retrieval and classification tasks  \cite{avni2011bovw, bouslimi2013bovw, meghana2017bovw_comparison}. In our study, we focused on the pathology-customized framework proposed by Zhu et al. \cite{zhu2018multiple}. Figure \ref{fig:BoVW} shows the general structure of BoVW indexing to generate a histogram as the index for patches or the entire WSI. 

\subsection{Yottixel} 

\begin{figure*}[htbp!]
    \centering
    \includegraphics[width=1\textwidth]{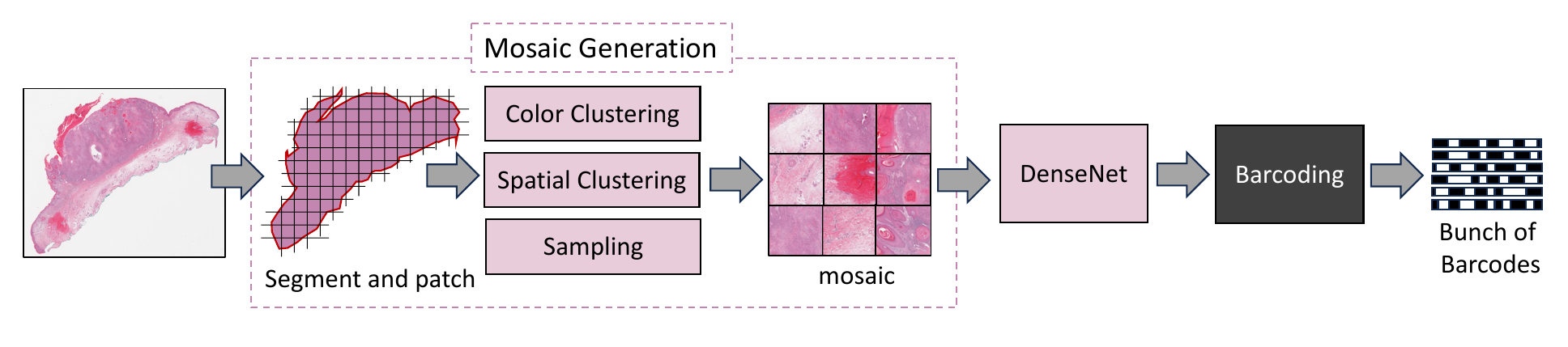}
    \caption{Yottixel indexing pipeline. Yottixel introduced two novel components: mosaic generation and barcoding. The mosaic is the result of a two-stage clustering that creates a small set of patches in an unsupervised fashion. Barcoding is the binarization of deep feature vectors to generate a \emph{bunch of barcodes}. Yottixel does not offer a trained network for feature extraction and uses a pre-trained network as a placeholder; any deep network can be used to provide deep features for the mosaic. }
    \label{fig:Yottixel}
\end{figure*}

Yottixel (a portmanteau for ``one yotta pixel'') is a search engine proposed for both patch and WSI indexing and search \cite{yottixel}. Yottixel introduced the first unsupervised patching method consisting of three stages: (1) cluster all patches based on their stain (color) histogram, 2) cluster patches additionally based on their proximity, 3) apply intra-cluster sampling at a desired rate. Yottixel patching generates a small set of patches called \emph{mosaic} to represent the WSI.  Subsequently, deep features can be extracted for each mosaic patch. Yottixel uses \emph{DenseNet} \cite{huang2017densely} as a placeholder such that any other deep feature extractor can be employed. Finally, Yottixel applied a binarization method, called \emph{MinMax} barcoding, or thresholding \cite{tizhoosh2015barcode,tizhoosh2016minmax,kumar2018deep}, an algorithm to approximate the 1D derivatives of deep features to generate a barcode for each deep feature vector from the mosaic, leading to a \emph{bunch of barcodes} (BoB) to represent the entire WSI. The BoB index enables efficient storage and fast search for similar WSIs. Figure \ref{fig:Yottixel} shows the structure of Yottixel indexing to generate a barcode for a single patch or a bunch of barcodes for the entire WSI. 

\subsection{SISH}

\begin{figure*}[htbp!]
    \centering
    \includegraphics[width=1\textwidth]{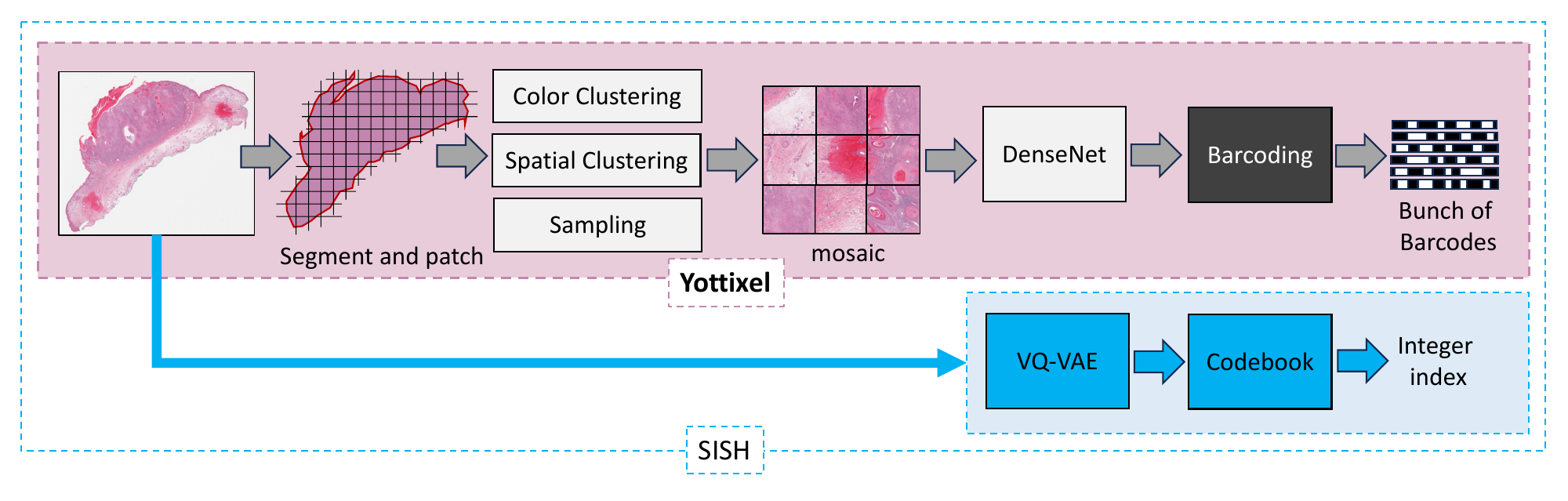}
    \caption{SISH indexing pipeline. SISH uses the entire Yottixel chain (pink part) and adds an autoencoder and codebook as an additional indexing scheme (blue part). SISH then uses a tree to match patches. As the authors report, this addition to Yottixel does not yield good results \cite{sish}, so they post-process the search results with a separate ranking scheme after the search.}
    \label{fig:SISH}
\end{figure*}

SISH (Self-supervised Image Search for Histology) \cite{sish} is another search pipeline. It uses Yottixel's indexing chain and adds an additional network on top of Yottixel's configuration.  It also adopts Van Emde Boas (vEB) tree \cite{vEBtree} for faster search. SISH indexes the images by using both Yottixe's barcodes and integers driven from an autoencoder, a vector quantized-variational autoEncoder (VQ-VAE). The authors of SISH report the attempt to improve Yottixel by adding the autoencoder and using integer indices via a tree fails to deliver better results. Hence, SISH uses a post-search ranking algorithm to improve the accuracy. Figure \ref{fig:SISH} shows the structure of SISH indexing.

\subsection{RetCCL}

\begin{figure*}[htbp!]
    \centering
    \includegraphics[width=0.8\textwidth]{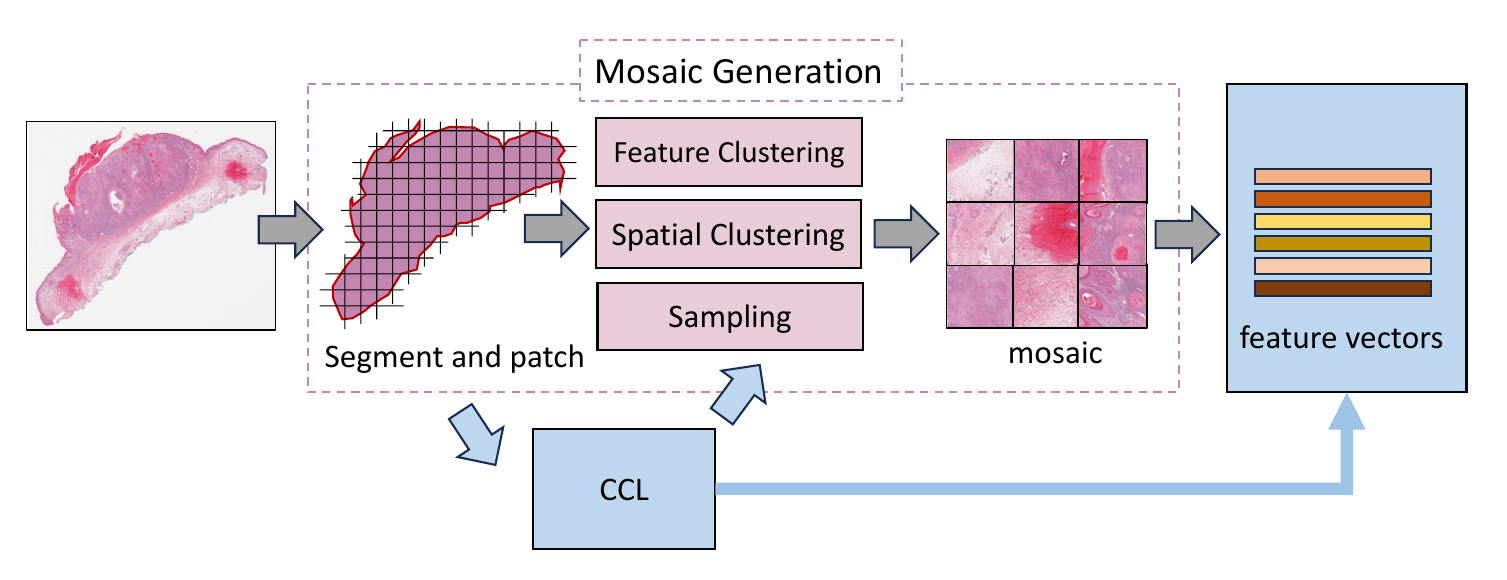}
    \caption{RetCCL indexing pipeline. RetCCL uses Yottixel's mosaic but replaces color histograms with deep features from the CCL network, a custom-trained network (blue blocks). It then uses cosine similarity to compare patches. RetCCL uses the post-search SISH ranking algorithm.}
    \label{fig:RetCCL}
\end{figure*}

The RetCCL (Retrieval with Clustering-guided Contrastive Learning) \cite{wang2023retccl} uses Yottixel's mosaic generation with one modification: it replaces color histogram with deep features. RetCCL is mainly concerned with the CCL deep network trained for histopathology. The mosaic patches are put through the CCL network to acquire deep feature vectors. Cosine similarity is then used to compare patches. RetCCL uses the SISH ranking algorithm outside of the search platform to improve its results. Figure \ref{fig:RetCCL} shows the indexing pipeline of RetCCL.

\section{Methods}
This section describes the process of conducting our experiments, how the search engines were implemented, what pre-trained networks were involved, and what datasets were used.

\subsection{Experimental Setup} \label{sec:expsetup}

We started with testing each of BoVW, Yottixel, SISH, and RetCCL on all four internal and three external datasets using their original configurations as implemented in their respective papers. It is crucial to externally verify the results of each of these search engines using their intended versions. Since there are similarities and differences in the algorithmic structure of these search engines, we also created some new variants of these search engines, e.g., with and without ranking. These studies help us understand the role of each part of the algorithm in the final performance of the search engine.

Of the four selected search engines, Yottixel and BoVW are not dependent on any specific deep network as the backbone. They can accommodate features from any deep network, but Yottixel uses a pre-trained DenseNet121 to extract deep features and BoVW (the implementation we studied) uses local binary patterns (LBP) as the feature. Generally, we can assume that the use of in-domain features (trained specifically for histology) results in much better tissue features. Since both SISH's autoencoder and the CCL-based network that RetCCL uses are trained on histopathology images, we replaced DenseNet in Yottixel with KimiaNet to verify the impact. KimiaNet is a trained network for histopathology image representation. It is a DenseNet-121 architecture, fine-tuned and trained with histopathology images from TCGA \cite{kimianet}. We also replaced the LBP features of BoVW with a customized autoencoder so all the search engines utilize deep features learned from tissue morphology.

One other feature shared by two of the search engines is ranking as an additional post-processing step. The post-search ranking is crucial for the effectiveness of SISH \cite{sish}. RetCCL also employs the same ranking algorithm \cite{wang2023retccl}. Since both of these search engines mention how important their ranking algorithms are to the success of their methods, we wanted to do testing where we removed the post-search ranking from SISH and RetCCL and also added it to Yottixel. These actions create one new variant for each one of SISH, RetCCL, and Yottixel. Finally, eight different search engines were selected for benchmarking, including the original and altered versions of the search engines: BoVW, Yottixel, Yottixel-K (using KimiaNet), Yottixel-KR (using KimiaNet and post-search ranking), SISH, SISH-N (no post-search ranking), RetCCL, RetCCL-N (no post-search ranking).

The benchmarking process for these search engines was executed to verify their search accuracy, speed, reliability, and storage requirements. This comprehensive evaluation provided a more holistic understanding of the strengths and limitations of these search engines, enabling a more objective comparison of their capabilities.

Search \textbf{accuracy} was measured among the top-1, majority vote among top-3, and majority vote among top-5 WSI matching results. The $F_1$ score was used as a measure to quantify both precision and recall. Additionally, the macro-averaged accuracy ($\bar{F}_1$) was calculated for each search engine in each experiment. All the experiments regarding SISH, Yottixel, and RetCCL were carried out using a \emph{leave-one-patient-out} approach which is the extreme form of cross-validation where the number of folds is equal to the number of patients in the dataset. Since the BoVW approach requires training an autoencoder for each dataset, the experiments for this specific search engine were carried out via a 5-fold cross-validation. We also measured the \textbf{speed} of each search engine in indexing and searching, separately. The average time taken to create the index was reported in minutes and the average search times (including both indexing and matching time for the query) were reported in seconds. To quantify \textbf{reliability}, we also counted how many times each search engine failed to process a WSI in each dataset. The failed WSIs were excluded from that specific experiment of search engine and dataset. We also measured the \textbf{storage need} of each search engine, separately. After indexing all the WSIs in each dataset, we measured the total disk space each search engine occupied to index the WSIs, excluding failed ones, to obtain the average disk space needed for each WSI. We also extrapolated the estimated amount of disk space needed to index one million WSIs using the estimated average. \textcolor{black}{The significance of storage is amplified by the desire for \textbf{democratization of AI} and the research community ambitions for inclusion and equity. WSIs being much larger  than radiology images have to be handled on top of ``already financially and operationally stressed healthcare system'' \cite{ ahmad2021artificial}. Most labs in developing countries do not have resources to store WSIs  \cite{zehra2023suggested}. Asking them for excessive extra storage to run a search engine appears illusory, a challenge that is widely recognized \cite{vuppalapati2021democratization}. }

All the experiments were executed on a Linux-based server with two AMD EPYC 7413 CPUs and 4 NVIDIA A100 GPUs with 80GB memory. We used one dedicated GPU
 for relevant processing steps 
 that required GPU (i.e., feature extractions), and CPU for matching. The environment was similar for all the search engines and all other non-system processes were terminated during the experiment to ensure the same computational power was available to each of them.

\subsection{Implementation}

\textbf{BoVW.} To build a bag of visual words, also known as visual dictionary, a variational autoencoder (VAE) was trained on 16x16 patches obtained at 20x magnification extracted from the WSIs in the training set. Subsequently, the test dataset was utilized to extract 4000 visual words from each WSI. The obtained deep features from the pre-trained VAE model were subjected to k-means clustering, utilizing 2048 clusters (the size of the dictionary), and computed across all deep features. Histograms of visual words were then computed for each WSI based on the resulting k-means clusters and the dictionary created. The collection of these histograms and their corresponding labels constitutes a search engine. The same indexing process is performed for query WSIs to obtain the histogram, utilizing the chi-squared metric, $\chi^2$, as the distance between the query WSI and any WSI already indexed in the dataset.

\textcolor{black}{We cannot claim to have trained the best autoencoder for BoVW. However, by introducing deep features into BoVW, we were able to include this well-known framework in the validation process.}

\textbf{Yottixel.} We used the sample Python code publicly published on KimaLab's website\footnote{https://kimialab.uwaterloo.ca/kimia/index.php/yottixel/}. The published code demonstrates the steps needed to index and search one sample WSI. We used the exact code to index all the WSIs iteratively. We then utilized the saved indexed images to search for a query WSI. The sample code uses a pre-trained DenseNet121 for feature extraction which is the original proposed version of Yottixel. As previously explained in the experiment setup section, we also tested two new variants of the original Yottixel. One variant was created by swapping the DenseNet121 with KimiaNet as the feature extractor. The other variant was implemented to use post-search ranking, as it is already implemented in SISH. To do so, we got the patch-level features extracted by Yottixel and applied the same ranking used by SISH. It is noteworthy that this change rendered Yottixel unable to match WSI-to-WSI, as this is already the case for SISH.

\textbf{SISH.} We were able to use the SISH code publicly published on its GitHub repository\footnote{https://GitHub.com/mahmoodlab/SISH}. The code is written in Python. There were minor adjustments such as updating all packages in their environment to work with our CUDA devices, changing the leave-one-out method to match the data we tested on, and also changing the evaluation metrics that are provided in their ranking algorithm to be consistent with the other search engines. 

\textbf{RetCCL.} As for RetCCL, only the CCL-based network is available on the GitHub repository\footnote{https://GitHub.com/Xiyue-Wang/RetCCL}. For image retrieval, the GitHub \emph{readme file} says ``Please refer to SISH, when clustering and searching, use our features, then remove the tree and search directly.'' Using the SISH code as guidance (although SISH uses the Yottixel's mosaic), we had to rewrite portions of the code, so the results may not be merely a duplicate of SISH. We used the patches that were generated by the SISH code (which is based on Yottixel's mosaic), and then extracted features using the RetCCL network. We followed the structure of the mosaic except used deep features instead of stain color for k-means clustering. Once we had mosaics for every WSI, we performed cosine similarity matching as proposed in the RetCCL paper. We made sure to have the results have the same format as the SISH so we could use the SISH ranking algorithm, as it is mentioned in the RetCCL paper that their ranking is inspired by SISH. The only difference is that we had to sort by cosine similarity instead of Hamming distance. 

The code that we used to recreate all of these search engines is available on the Rhazes Lab's GitHub page\footnote{https://github.com/RhazesLab}.

\subsection{Pre-Trained Networks}

Besides the deep networks published as part of the search engine methods, \textcolor{black}{we used two other pre-trained networks as the backbone, as needed. One is DenseNet121 \cite{denesenet_huang2018} and the second one is 
KimiaNet  \cite{kimianet}.}

\subsection{Datasets}

We conducted our benchmarking on four internal and three public datasets. The internal datasets are collected from the Mayo Clinic each from a different organ with different labels: breast epithelial tumor subtypes, fatty liver disease, cutaneous squamous cell carcinoma (cSCC) subtypes, and colorectal polyps. Table \ref{tab:Datasets} gives a detailed overview of the number of patients, WSIs, and classes in these public/internal datasets. 

\begin{table*}[htb]
    \centering
    \caption{Validation Datasets. The number of patients, WSIs, and classes in each dataset used for benchmarking the search engines.}
    \label{tab:Datasets}
    \begin{threeparttable}
    \begin{tabular}{clcccc}
        \toprule
        \multirow{4}{*}{\rotatebox[origin=c]{90}{Internal}}& Dataset & Number of Patients & Number of WSIs &  Number of Classes\\
        \cline{2-5}
        &Breast Epithelial Tumors & 74 & 74 & 16 + normal\\
        \cline{2-5}
        &Fatty Liver Disease & 326 & 326 & 2 + normal \\
        \cline{2-5}
        &cSCC & 660 & 660 & 3 + normal \\
        \cline{2-5}
        &Colorectal Polyps & 209 & 209 & 3 \\
        \hline\hline
        \multirow{3}{*}{\rotatebox[origin=c]{90}{Public}}&CAM16 \cite{DataCAMELYON16} & 129 & 129 & 2 \\
        \cline{2-5}
        &BRACS \cite{DataBRACS} & 87 & 87 & 3 \\
        \cline{2-5}
        &PANDA \cite{DataPANDA} & 1000 & 1000 & 6 \\
        \bottomrule
    \end{tabular}
    \begin{tablenotes}\footnotesize
        \item cSCC = cutaneous squamous cell carcinoma
    \end{tablenotes}
    \end{threeparttable}
\end{table*}

\subsubsection{Internal Datasets}
\begin{figure*}[htb]
    \centering
    \includegraphics[width=\textwidth]{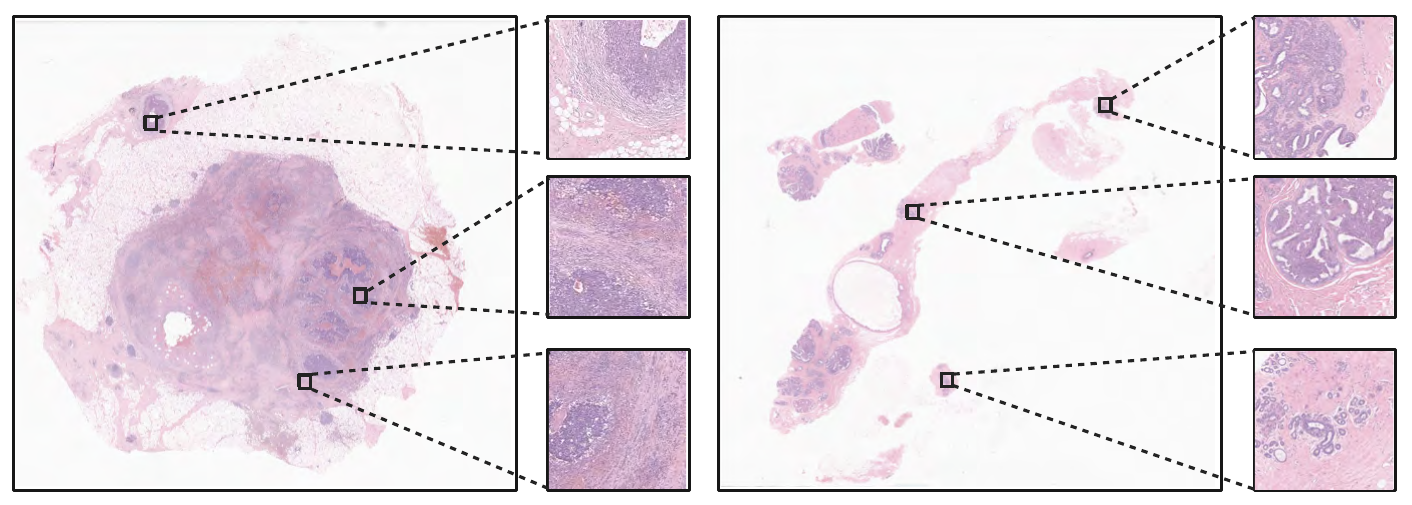}
    \caption{Thumbnails of two sample WSIs from the breast dataset:  (left) invasive breast carcinoma of no special type, (right) intraductal papilloma with columnar cell Lesions.}
    \label{fig:breast_examples}
\end{figure*}

\textbf{Breast Epithelial Tumors Dataset.} This dataset includes samples from 74 patients with 16 histologic subtypes of breast epithelial tumors (many rare subtypes) including adenoid cystic carcinoma, adenomyoepithelioma, ductal carcinoma in situ, ductal carcinoma in situ with columnar cell lesions including flat epithelial atypia, atypical ductal hyperplasia, intraductal papilloma with columnar cell lesions, invasive breast carcinoma of no special type, invasive lobular carcinoma, lobular carcinoma in situ with atypical lobular hyperplasia, lobular carcinoma in situ with flat epithelial atypia and atypical lobular hyperplasia, malignant adenomyoepithelioma, metaplastic carcinoma, microglandular adenosis, micro-invasive carcinoma, mucinous cystadenocarcinoma, and radial scar complex sclerosing lesion. This dataset also included cases of normal breast tissue. Some classes included only one sample which has rendered this dataset unsuitable for a majority vote search approach. Figure \ref{fig:breast_examples} shows the thumbnails of two sample WSIs of the breast dataset along with random sample patches.

\begin{figure*}[htb]
    \centering
    \includegraphics[width=\textwidth]{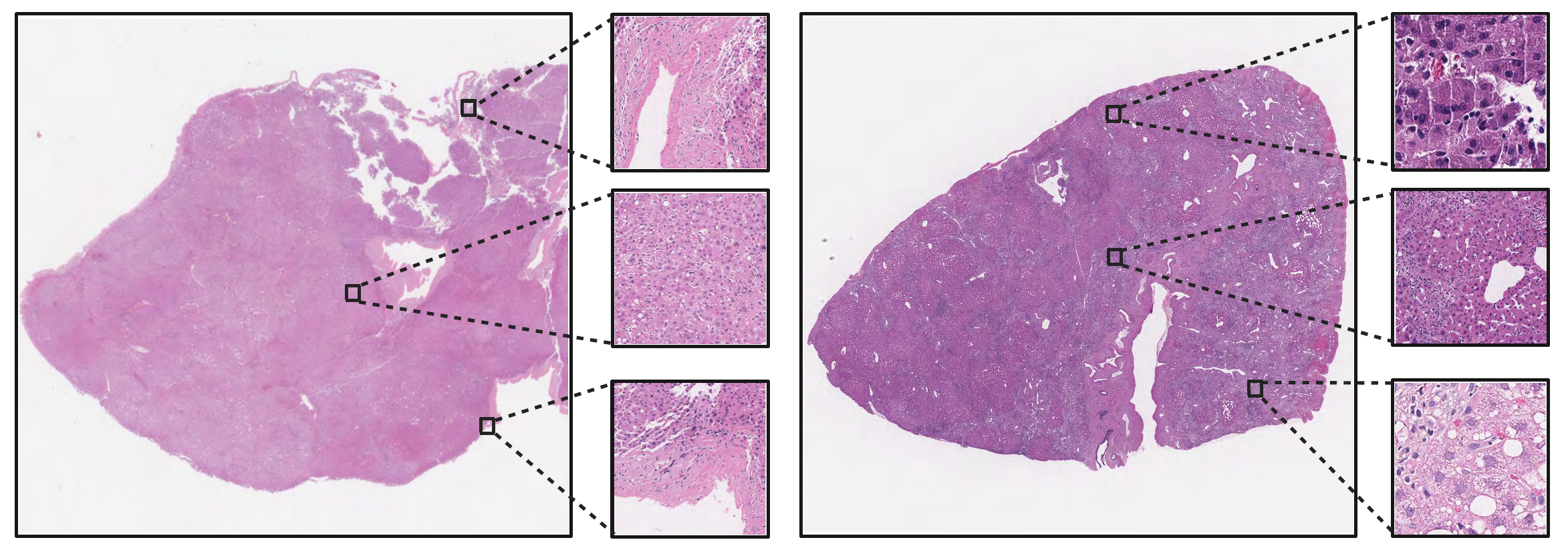}
    \caption{Thumbnails of two sample WSIs from the liver dataset:  (left) NASH, (right) ASH.}
    \label{fig:liver_examples}
\end{figure*}

\textbf{Fatty Liver Disease Dataset.} This dataset includes 326 liver biopsy slides obtained from patients diagnosed with either alcoholic steatohepatitis (ASH) or non-alcoholic steatohepatitis (NASH). This dataset also includes WSIs of normal liver tissue. The diagnosis of ASH was made based on the chart review and expert opinion on history, clinical presentation, and laboratory studies. Liver biopsies from a cohort of morbidly obese patients undergoing bariatric surgery were used to select cases for the NASH group. In total, 150 WSIs of patients with ASH, 158 WSIs of patients with NASH, and 18 WSIs of normal individuals were included. Figure \ref{fig:liver_examples} shows the thumbnails of two sample WSIs of the liver dataset and random sample patches.

\begin{figure*}[htb]
    \centering
    \includegraphics[width=\textwidth]{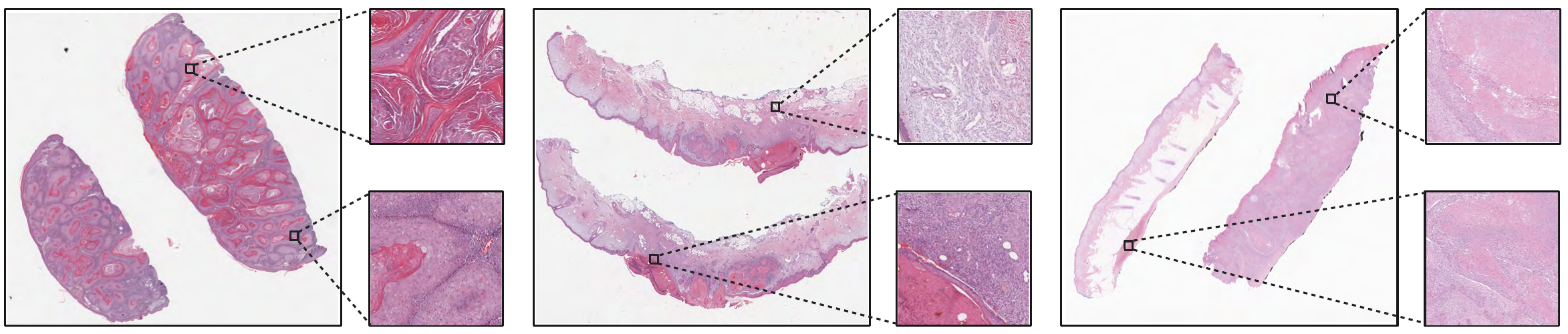}
    \caption{Thumbnails of three sample WSIs from the skin dataset:  (left) well-differentiated, (middle) moderately differentiated,  (right) poorly differentiated cSCC.}
    \label{fig:skin_examples}
\end{figure*}

\textbf{Cutaneous Squamous Cell Carcinoma Dataset.} The skin dataset included a total number of 660 skin tissue WSIs of patients diagnosed with cutaneous squamous cell carcinoma (cSCC). The data was pulled from the internal Mayo Clinic database (REDCap) including 386 slides of well-differentiated, 100 moderately differentiated, and 67 poorly differentiated cSCC. There were also 107 normal slides selected to represent the normal skin tissue as a separate group. Figure \ref{fig:skin_examples} shows the thumbnails of three sample WSIs from this dataset and random sample patches.

\begin{figure*}[htb]
    \centering
    \includegraphics[width=\textwidth]{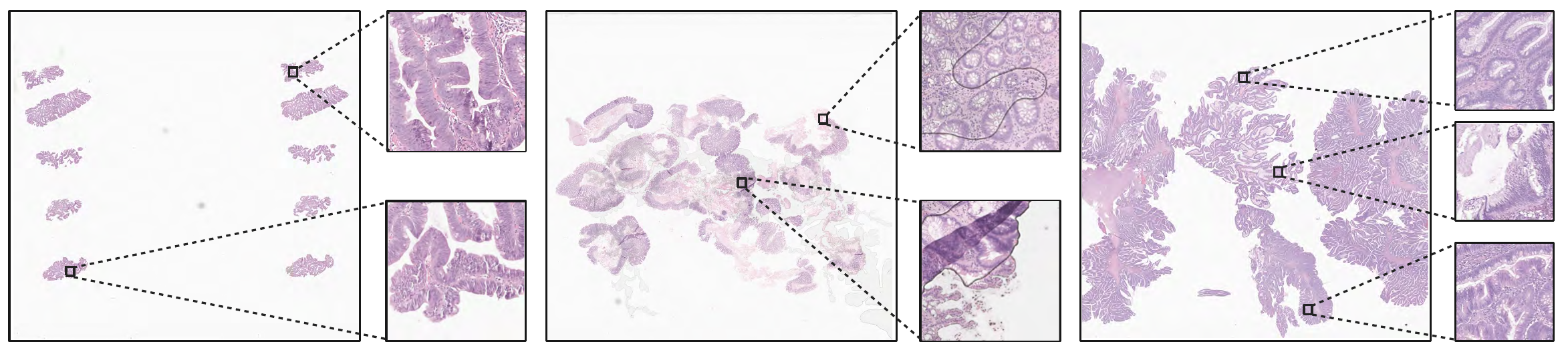}
    \caption{Thumbnails of three sample WSIs from the CRC dataset:  (left) recurrent polyp, (middle) cancer adjacent polyp, (right) non-recurrent polyp.}
    \label{fig:crc_examples}
\end{figure*}

\textbf{Colorectal Polyps Dataset.} The colorectal (CRC) dataset was also collected at Mayo Clinic and comprises a total of 209 WSIs. This dataset included three distinct categories established for colorectal pathology, including 63 slides of cancer adjacent polyps (CAP), 63 slides of non-recurrent polyps (POP-NR), and 83 slides of recurrent polyps (POP-R). This dataset had no normal tissue included. Figure \ref{fig:crc_examples} shows the thumbnails of three sample WSIs from this dataset along with sample random patches.

\subsubsection{Public Datasets}
Although we are convinced that measuring the search performance on high-quality clinical data is the ultimate performance validation for image search technologies, we do see value in confirming the internal results by repeating the experiments for public datasets. As TCGA, as the largest public dataset has been used for training and testing of many available solutions, we did select the following well-known datasets to repeat our experiments. 

\textbf{PANDA - Prostate Dataset \cite{DataPANDA}}. This is an open-access repository comprising $12,625$ WSIs depicting prostate biopsies, stained with H\&E, meticulously gathered from varied international sources to facilitate a thorough assessment of methods for prostate cancer detection and grading. It contains 6 ISUP-grading from 0 to 5 which are used as classes. Since the only available set is the training set with $10,616$ WSIs, we picked the first $1,000$ WSIs for our leave-one-patient-out validation.

\textbf{CAMELYON16 Breast Dataset \cite{DataCAMELYON16}}. This well-known dataset, short \textbf{CAM16}, contains a set of $399$ diligently annotated WSIs representing lymph node sections derived from breast cancer patients at two healthcare facilities in the Netherlands. It contains two classes Tumor and Normal. We used the testing subset of $129$ WSIs.

\textbf{BRACS Breast Dataset \cite{DataBRACS}}. This dataset contains $547$ WSIs obtained from $189$ patients, meticulously annotated into seven distinct lesion subtypes by board-certified pathologists. We used the testing set which contains $78$ WSIs with three classes: Benign, Atypical, and Malignant.

\vspace{0.1in}

\textcolor{black}{Working on rather small datasets with a limited number of classes/labels may not fully replicate the real-world application of image search if applied on larger WSI repositories. However, large WSI collections on high-performance storage devices with access to GPU computation does not seem to be available yet.}

\section{\textcolor{black}{Analysis and} Results}
\textcolor{black}{We first conduct \textbf{analysis} of the search engines} in terms of their algorithmic structure, search capabilities, training, and test data used. Subsequently, we conducted extensive \textbf{experiments} to compare their search performance. 

\textcolor{black}{\textbf{Analysis --}} As for algorithmic structure comparison, we compared the patching methods they used to break down the WSI and perform indexing and matching. Overall, by comparing their search capabilities, we determined whether they are able to process/search for both patch and WSI matching. Training and testing data comparisons encompass a detailed inventory of the data each search engine was trained with or validated on. 

\textcolor{black}{\textbf{Experiments --}} We used both high-quality clinical datasets and well-known public datasets to quantify the search performance of each method. We examined accuracy, time measurements for indexing and search, robustness, and storage requirements as main performance benchmarking. We standardized the implementation by utilizing a consistent computational machine and refrained from employing multiprocessing techniques, ensuring equitable time readings for all methods.

\subsection{\textcolor{black}{Analysis of} Algorithmic Structure}

Table \ref{tab:Algorithmic Comparison} offers an overview of the algorithmic structures employed by the four search frameworks (also see Figures \ref{fig:BoVW},\ref{fig:Yottixel},\ref{fig:SISH} and \ref{fig:RetCCL} that visualize only the first three columns of Table \ref{tab:Algorithmic Comparison}). Each approach is detailed across six key stages: Divide, Features, Encoding, Space, Matching, Speed, and Post-Processing. To analyze the algorithmic structure, we considered both the descriptions in the corresponding paper, as well as details of their implementation. We did not list ``pre-processing'' as all four methods start with segmenting the slide into tissue and non-tissue regions.

\begin{table*}[htb]
\centering
\begin{threeparttable}
    \centering
    \caption{Algorithmic comparisons of all search methods (also see Figures \ref{fig:BoVW},\ref{fig:Yottixel}, \ref{fig:SISH} and \ref{fig:RetCCL})}
    \label{tab:Algorithmic Comparison}
    \scriptsize 
    \begin{tabular}{lcccccc|c}
        \toprule
        
        \makecell{} &  Divide & Features   & Encoding & Space  & Matching & Speed & Post-Proc.\\
        \hline
        BoVW \cite{zhu2018multiple} & \ \makecell{Sampling} & \makecell{Autoencoder\tnote{1}} & Clustering & $\mathcal{O}(N)$ & \makecell{Euclidean\\Distance} & $\mathcal{O}(N)$ & None \\
        \hline
        Yottixel \cite{yottixel} &  \makecell{Mosaic} & \makecell{DenseNet} & Barcoding & $\mathcal{O}(N)$ & \makecell{Hamming \\Distance} & $\mathcal{O}(N)$ & None \\
        \hline
        SISH \cite{sish}&  \makecell{Yottixel's \\Mosaic\tnote{2}} & \makecell{DenseNet \\+ VQ-VAE}  &  \makecell{Yottixel's \\Barcoding \\+ Long integers} & $ \mathcal{O}(2^m)$ & \makecell{Hamming \\Distance \\+vEB Tree} &  $\mathcal{O}(\lg \lg N)$ & Re-Ranking \\
        \hline
        RetCCL \cite{wang2023retccl}&  \makecell{Yottixel's\\ Mosaic\tnote{3}} & CCL\tnote{4} & None & $\mathcal{O}(N)$ & \makecell{Cosine \\Similarity} & $\mathcal{O}(N)$ & Re-Ranking \\
        \bottomrule
    \end{tabular}
    \begin{tablenotes}\footnotesize
        \item[1] In the original paper, BoVW uses local binary patterns (LBP) as features. We replaced it with an autoencoder to adjust the experimental comparison to the deep-learning level for all search methods.       
        \item[2] SISH uses Yottixel's mosaic with no modification.
        \item[3] RetCCL uses Yottixel's mosaic but instead of color histogram, it uses deep features for all patches.
         \item[4] RetCCL extracts features for all patches prior to the mosaic generation.
    \end{tablenotes}
    \end{threeparttable}
\end{table*}

\textbf{Divide --} Since using all WSI patches for the downstream tasks is inefficient in terms of both computation and storage (as this is the case for SMILY \cite{hegde2019similar}), search methods must employ techniques to select a subset of the patches, also known as the ``divide'' stage in a \emph{divide \& conquer} attempt to master the gigapixel nature of WSIs \cite{tizhoosh2024image}. While BoVW uses a random sampling approach to select visual words (which are much smaller than patches), Yottixel uses its ``mosaic'' approach, an unsupervised approach to get a small set of patches. SISH does not offer any new solution and uses Yottixel's mosaic, the same as RetCCL. Yottixel performs color and spatial clustering to select representative patches, whereas RetCCL uses deep feature and spatial clustering, a much more expensive variation.

\textbf{Features --} Every search method has to utilize deep neural networks to obtain deep features from the image data, i.e., patches. RetCCL uses its clustering-guided contrastive learning (CCL) trained network using TCGA images to extract features for all patches prior to the mosaic generation step since it utilizes them for the divide step. In contrast, Yottixel and SISH both generate feature vectors only for the mosaic patches. Yottixel uses a pre-trained DenseNet, which can be swapped by any other network. SISH also extracts features using DenseNet, with the addition of an autoencoder trained on TCGA images. The rationale behind integrating the autoencoder alongside DenseNet in SISH is not sufficiently justified. It has been argued that SISH does not use any self-supervision for training the autoencoder, hence the paper may have introduced a misnomer  \cite{sikaroudi2023comments}. Originally, BoVW used local binary patterns (LBP) as features but for the sake of comparison, we replaced it with an autoencoder so all the search engines utilize deep features.

\textbf{Encoding --} The three search engines besides RetCCL employ some kind of encoding on the deep features to improve the search speed and/or reduce storage needs. BoVW utilizes clustering to obtain the ``histogram of visual words'' for the WSI. Yottixel converts the deep features into binary codes, i.e., barcodes, using a MinMax algorithm \cite{tizhoosh2015barcode,kumar2018deep,yottixel}. SISH uses Yottixel's barcodes (although corresponding papers have not been cited) but also creates an integer value for each patch by putting the discrete latent code through a pipeline of average pooling, shift, and summation algorithms. The addition of these integer codes is motivated by the desire for constant speed through the usage of a tree but any benefit is lost when more and more images are indexed due to double-indexing and massive overhead of the specific tree used. RetCCL, in contrast to the other three methods, has no encoding technique.

\textbf{Space (storage need) --} The encoding has a cost: memory to store the index (i.e., features). As the Divide step splits the WSI into $N$ patches, one can talk about \emph{theoretical} upper bound (the so-called big-Oh notation) for storage. BoVW, Yottixel, and RetCCL have a linear upper bound, $\mathcal{O}(N)$, for storage. That means their need for storage increases linearly as the number of patches $N$ increases. Of course, this is just a rough estimate: BoVW stores one vector of integers for WSI, Yottixel saves 70-100 binary vectors \cite{yottixel} for the mosaic, and RetCCL saves 70-100 real-values vectors for the mosaic. These three indexes will not be the same. SISH is an exception: Using a special tree called vEB tree \cite{van1975preserving} forces SISH to save very long integers in the encoding space $m$, and not in the patch space $N$. The high speed of the tree comes with a catch: the space requirement is exponential, $\mathcal{O}(2^m)$. Although SISH authors mention this, they state that  ``$m$ remains fixed and does not scale with the number of data points'' but this is not, as our experiments showed, a remedy for the excessive storage need. The storage need can be reduced to $\mathcal{O}(N)$, e.g., through multiple hashing tables at all tree levels, and at the expense of code complexities, but the excessive overhead remains. The unitless number of $2^m = 1,125,899,906,842,624$ that the paper suggests is the order of magnitude that results in terabytes of storage even for small datastes as we will report. The maintenance of such tree structures is another factor that would render the usage of such data structures infeasible.      

\textbf{Matching (or Search) --}    
In order to find the best matches, every search method utilizes a means of comparing the indexed images. BoVW can use any distance metric for histogram matching to measure image similarity although $\chi^2$-distance is a preferred comparison metric for histograms. Yottixel employs a simple but effective approach to image-to-image comparison via \emph{median of minimum} Hamming distances among the barcodes of all mosaic patches of two WSIs.  SISH uses a vEB tree to find the best candidates, as well as Hamming distance of patch barcodes. RetCCL calculates cosine similarity between patch feature vectors. All search methods have a \emph{theoretical} upper bound of $\mathcal{O}(N)$, except SISH that claims to perform search in \emph{constant time} because of the upper bound $\mathcal{O}(\lg \lg N)$ (for instance, for one billion patches, roughly 10 million WSIs, we have $\lg \lg (1,000,000,000) \approx \lg (30) \approx 5 $, which means we can search in 10 million WSIs in 5 calculations which would a fraction of a second. Unfortunately, as the space analysis shows, this is just theory. For such a large number of images, SISH would collapse under the burden of terabytes of memory needed by the tree. The vEB tree is rather an outdated data structure and some of its aspects are inspiring for other data structures like B-trees \cite{bender2005cache} but it cannot be used in the practice of hyperdimensional applications. This is why it has been mostly used as a priority queue for non-vision applications with an array of bare numbers under massive parallel processing (e.g., 10 processors for 1 GB encoding space)\cite{wang2007pipelined}, and not for very long integers representing complex tissue. Hence, this is not surprising that CBIR literature has never reported the usage of vEB trees.  

We expect that BoVW with one histogram per WSI may be the fastest approach, followed by Yottixel due to using several binary vectors. Due to their lean memory requirements, the implementation of both BoVW and Yottixel can be optimized to achieve  $\mathcal{O}(\log N)$.

\textcolor{black}{We should bear in mind that the Big-Oh notation is a theoretical upper bound with some practical limitations. For instance, comparing $N$ patches with $M$ other patches is bounded by $\mathcal{O}(N^2)$ for $N\!>\!M$. However, if we use binary features (like Yottixel's barcodes) and measure similarity via Hamming distance, then the actual time is rather bounded by the  $\mathcal{\theta}(N^\alpha)$ with $\alpha \in (0,1]$.}

\textbf{Post-Processing --} Generally, the search process concludes once results have been retrieved and sorted based on their similarity to the query. However, both SISH and RetCCL utilize an additional ranking algorithm after the search, which eliminates patches deemed less likely to be similar to the query. \textcolor{black}{The intuition behind ranking is to find the most promising patches coming from the search engine on the basis of the uncertainty \cite{sish}.  It relies on three helper functions \underline{Weighted Uncertainty}  (the uncertainty for each patch by computing their entropy), a \underline{Clean} function (to remove outliers and the patches that are less similar to the query) and a \underline{Filtered-By-Prediction} function (to remove patches with low likelihoods).} 

The inclusion of ranking as a post-processing step is evidently motivated by a lack of confidence in the search results. This approach seems misguided, as relying on post-search ranking implies a lack of trust in the initial results. The authors of SISH stress the significance of ranking for their outcomes, indicating that incorporating an autoencoder into the Yottixel pipeline has not yielded any discernible benefits, despite the paper emphasizing the autoencoder's role.

\subsection{\textcolor{black}{Analysis of} Search Capabilities}

Processing a large number of patches is expected from a histopathology search engine. More importantly, a search engine must be able to process WSIs (break them into a set of patches) for both patch-to-patch and WSI-to-WSI searches. Therefore, each search engine's capabilities in search can be broken down into two main categories: ``Patch Processing" and ``Patch Search," which pertain to the processing of image patches, and ``WSI Processing" and ``WSI Search," which relate to whole slide image processing and searching. 

It is an indispensable capability of a search engine in histopathology to be able to perform both patch search and WSI search. \textcolor{black}{WSI-to-WSI matching compares representative WSI patches to representative WSI patches in absence of ROI delineations. These representative WSI matches, like Yottixel’s mosaic, are expected to be the result of some unsupervised patch selection method. The query, coming from the pathologist, could be the entire WSI or selected ROI but the indexed WSIs are generally not expected to have ROIs, especially for large repositories. Comprehensive WSI-to-WSI comparison is important as several malignancies, or multiple staging/grading quantities may be present in the image (see the \textcolor{red}{Fig. S1 in Supplementary Materials} that illustrates a simple case of prostate Gleason Scoring that may be falsified if we remove some of representative WSI patches).}

Clearly, WSI-2-WSI matching (WSI search) is a more difficult task compared to patch-to-patch matching. Also, WSI search is pivotal for \textcolor{black}{patient tissue comparison}. Both BoVW and Yottixel stand out as comprehensive solutions with the ability to perform both patch and WSI searches. In contrast, SISH and RetCCL, while being able to process and search patches due to using the Yottixel's mosaic, lack the capability to perform WSI-2-WSI matching. \textcolor{black}{This in fact stems from their dependence on ranking patches after the search to find the label for the query WSI; ranking may separate the mosaic patches (some of them get high and some other low ranks). Therefore, although the label found by SISH and RetCCL may be correct based on top-$n$ patches, this does not constitute true WSI-2-WSI matching as the decision is not based on all mosaic patches}. In the case of SISH, even without post-search ranking, the tree used for fast processing cannot be restricted to top-$n$ WSIs.

\begin{figure*}[htb]
    \centering
    \includegraphics[width=\textwidth]{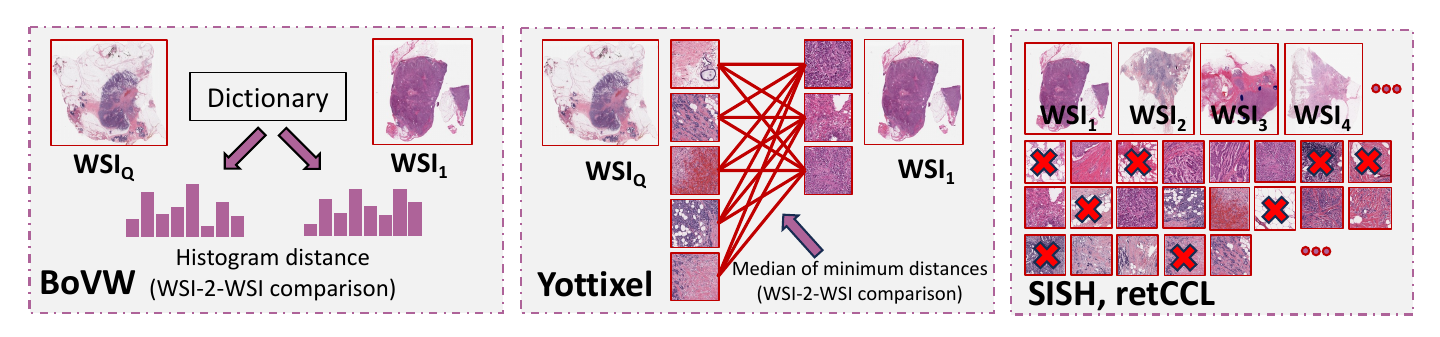}
    \caption{Both BoVW and Yottixel can perform WSI-2-WSI matching. \textbf{Left:} BoVW creates a histogram for an image (either a patch or a WSI). \textbf{Middle:} Yottixel creates a mosaic of patches for each WSI and uses the median of minimum Hamming distances to get a single dissimilarity value for WSI comparison. \textbf{Right:} The ranking that SISH and RetCCL use to achieve higher accuracy eliminates the possibility of a  WSI-2-WSI comparison and reduces the search to classification. In the case of SISH, the pre-ranking results delivered by the vEB tree it uses are patch-oriented as well. }
    \label{fig:rankingeffect}
\end{figure*}

In summary,   BoVW creates a histogram for an image (either a patch or a WSI). The histogram can be calculated for both patch and WSI (Figure \ref{fig:rankingeffect}, left). Hence, BoVW can process and match both patches and WSIs. Yottixel creates a mosaic of patches for each WSI and uses the ``median of minimum distances'' to get a single distance metric for WSI comparison (Figure \ref{fig:rankingeffect}, middle); Hence, Yottixel can match both patches and WSIs. Both SISH and RetCCL use Yottixel's mosaic but their ranking algorithms essentially force them to perform patch matching although the ranking may be configured to select one patch per WSI in order to emulate WSI selection. They take a query patch and compare it with every other patch in every other WSI (Figures \ref{fig:rankingeffect}, right). SISH pre-raking tree processing is not WSI-conform either. Hence, \emph{SISH and RetCCL are rather \underline{patch classifiers}} when using post-search ranking, hence not usable for \textcolor{black}{patient tissue comparison} in a clinical setting.

\subsection{\textcolor{black}{Analysis of} Training/Test Data}

The data used for any type of training, the test data, and the user feedback experiments are all important factors in evaluating search engines. Table \ref{tab:DataComp} offers an overview of the data utilized by the four search engines. \underline{BoVW} uses Kimia Path24\footnote{\url{https://github.com/RhazesLab/kimia_path24}} as testing data, comprising 24 different anatomic sites \cite{shafiei2021kimia_path24}. Notably, BoVW, as reported in \cite{zhu2018multiple}, does not use any deep features and does not conduct user tests. \underline{Yottixel} uses a pre-trained model, namely DenseNet. Hence, it has been validated on entire TCGA images, both diagnostic slides (FFPE slides) and frozen sections, in addition to some private data from UPMC, accompanied by some user tests involving three pathologists. The data Yottixle has been validated on covers a wide range of anatomic sites, totaling 27 primary sites, and an extensive number of subtypes. It is notable that since the pre-trained DenseNet121 has been trained on ImageNet, which is a non-pathological dataset, all Yottixel tests can be considered external validation. \underline{SISH} draws its training data from TCGA and conducts testing with a diverse dataset comprising TCGA, CPTAC, and BWH data, without involving user tests. It focuses on a reduced set of anatomic sites (13) but accounts for 56 different subtypes. It is not clear how the training and testing TCGA data for SISH have been separated. \underline{RetCCL} uses TCGA and PAIP data for training and only TCGA data for testing, without user tests. RetCCL considers 13 anatomic sites but focuses on a more limited set of subtypes. 

\begin{table}[htb]
    \centering
    \caption{\textcolor{black}{Comparison of Training and Test Data that each Search Engine has reported in corresponding papers.}}
    \label{tab:DataComp}
    \begin{tabular}{lllc}
        \toprule
         & Training Data & Test Data  & Users  \\
        \toprule
        BoVW & none  & Kimia Path24  & none \\
        \hline
        Yottixel& none & TCGA (frozen) & 3 \\
          &  & TCGA (FFPE) &     \\
        &  & UPMC &     \\
        \hline
        SISH & TCGA (FFPE) & TCGA (FFPE) & none \\
          &  & CPTAC, BWH &   \\
        \hline
        RetCCL & TCGA (FFPE), PAIP & TCGA (FFPE) & none \\
        \bottomrule
    \end{tabular}
\end{table}


\subsection{Comparative Experiments for Internal Data}

In this section, we report comprehensive results for search accuracy, indexing and search speed, reliability, and storage requirements of all four search engines on four internal and three public datasets. The results are summarized by also getting an overall performance rating $R^*$ for each methodology based on the individual ratings $R_1,R_2,\dots,R_k $ delivered by $k$ separate experiments:

\begin{equation}
    R^* = \frac{1}{k}\sum\limits_{i=1}^{k} R_i
\end{equation}

The rating $R_i$ for any method will simply be its position in a sorted list (proper order) based on its performance. \textcolor{black}{We consistently use the F1-score of top-$n$, or the majority of top-$n$, for accuracy calculation based on \emph{leave-one-WSI-out} validation. The F1-score is the harmonic mean of precision (number of the retrieved images actually relevant to the query (i.e., low false-positive rate) and recall (\textcolor{red}{number of images with the same class label as the query were actually retrieved,} i.e., low false-negative rate), providing a balanced measure of both metrics  \cite{tizhoosh2024image}. } 

\subsubsection{Search Accuracy Results}

Search accuracy is measured via $F_1$ Score (the harmonic mean of precision and recall). 
We measure the $F_1$ score for the top-1 search results, the majority vote among the top 3 search results (MV3), and the majority vote among the top 5 search results (MV5). Table \ref{tab:accuracy} presents a comprehensive comparison of the $F_1$ for search accuracy for each search engine across four internal datasets Liver, Skin, CRC (Colorectal Cancer), and Breast. The Breast dataset was excluded from the MV3 and MV5 search since there are classes in this dataset with as low as one sample. We also test other variants of SISH and RetCCL (with and without post-search ranking), and Yottixel (with DenseNet, and with KimiaNet with/without post-search ranking) to provide a more comprehensive comparison.

\begin{table*}[htbp!]
    \centering
    \caption{Average $F_1$ scores of  top-1 accuracy (upper section), majority vote among top-3, MV@3,  (middle section), and majority vote among top-5, MV@5,  (lower section) on internal Mayo Clinic's datasets. The best results are highlighted in green, the second best in blue colors.}
    \label{tab:accuracy}
    \begin{threeparttable}
    \begin{tabular}{lcccc|cc}
        \toprule
         & Liver & Skin & CRC & Breast & $\bar{F}_1$ & $R_{top1}$\\
                \midrule
        BoVW & 50.71$\pm$25.46 & 54.57$\pm$33.47 & 43.07$\pm$4.70 & 37.06$\pm$41.60 & 46.35$\pm$6.78 & 6\\
        \midrule
        Yottixel & \cellcolor{myblue}72.29$\pm$3.07 & 64.60$\pm$30.87& 57.19$\pm$19.81 & \cellcolor{myblue}38.05$\pm$38.34 & 58.19$\pm$14.88 & 3\\
        \midrule
        SISH & 71.18$\pm$10.43 & 69.40$\pm$7.61 & 56.51$\pm$26.28 & 17.02$\pm$31.68& 53.53$\pm$25.20 & 4\\
        \midrule
        RetCCL & 38.50$\pm$20.17 & 43.04$\pm$28.72 & 36.22$\pm$5.05 & 1.43$\pm$3.78 & 29.80$\pm$19.12 & 7\\
\midrule
        Yottixel-K\tnote{1} & 61.54$\pm$27.33& \cellcolor{myblue}69.73$\pm$29.55 & \cellcolor{myblue}59.91$\pm$15.01&\cellcolor{mygreen} 55.90$\pm$29.49 & \cellcolor{mygreen}61.77$\pm$5.81 & 1\\
        \midrule 
        Yottixel-KR\tnote{2} & \cellcolor{mygreen} 76.00$\pm$12.09 & \cellcolor{mygreen}70.67$\pm$29.84 & \cellcolor{mygreen}62.47$\pm$16.23 & 31.01$\pm$31.00 &  \cellcolor{myblue}60.04$\pm$20.14 & 2\\
        \midrule 
        SISH-N\tnote{3} & 63.70 $\pm$16.03 & 62.35$\pm$31.15 & 49.40$\pm$13.82 & 13.64$\pm$24.24& 47.27$\pm$23.33 & 5\\
        \midrule
        RetCCL-N\tnote{4} & 43.54$\pm$19.44 & 34.71$\pm$25.40 & 36.72$\pm$4.98 & 2.50$\pm$9.68 & 29.37$\pm$18.31 & 8\\
        \bottomrule
        \midrule
        & Liver & Skin & CRC & Breast & $\bar{F}_1$ & $R_{MV3}$\\ \hline
        BoVW & 54.54$\pm$29.49 & 57.26$\pm$36.37 & 58.18$\pm$4.21 & N/A & 56.66$\pm$1.55 & 6\\
        \midrule
        Yottixel & \cellcolor{mygreen} 76.89$\pm$1.34  & 66.56$\pm$34.88 & 59.18$\pm$ 21.62 & N/A & \cellcolor{myblue} 67.54$\pm$8.90 & 2 \\
        \midrule
        SISH & 69.58$\pm$ 14.97& \cellcolor{mygreen} 70.81$\pm$26.97 & 56.98$\pm$26.63 & N/A & 65.79$\pm$7.65 & 4\\
        \midrule
        RetCCL & 36.91$\pm$32.27 & 41.55$\pm$36.03 & 39.55$\pm$3.62 & N/A & 39.34$\pm$2.33 & 7\\
        \midrule 
        Yottixel-K\tnote{1} & 66.99$\pm$ 23.38& 70.09$\pm$32.23 & \cellcolor{mygreen}60.56$\pm$ 11.97 & N/A & 65.88$\pm$4.86 & 3\\%
        \midrule 
        Yottixel-KR\tnote{2} & \cellcolor{myblue}74.72$\pm$16.21 & \cellcolor{myblue}70.10$\pm$31.30 & \cellcolor{myblue}59.33 $\pm$19.13& N/A & \cellcolor{mygreen}68.05$\pm$7.90 & 1\\%
        \midrule 
        SISH-N\tnote{3} & 56.90$\pm$31.13 & 65.01$\pm$32.06& 53.61$\pm$16.01 & N/A & 58.51$\pm$5.87 & 5\\
        \midrule
        RetCCL-N\tnote{4} & 48.67$\pm$14.83 & 33.76$\pm$29.55 & 35.51$\pm$7.21 & N/A & 39.31$\pm$8.15 & 8 \\
        \bottomrule
                \toprule
         & Liver & Skin & CRC & Breast & $\bar{F}_1$ & $R_{MV5}$\\
                \midrule
        BoVW & 52.27$\pm$31.13 & 58.18$\pm$ 37.75& 43.85$\pm$4.21 & N/A & 51.43$\pm$5.88 & 6 \\
        \midrule
        Yottixel & \cellcolor{mygreen}77.33$\pm$5.03  & 66.52$\pm$35.95& \cellcolor{mygreen}65.44$\pm$18.77 & N/A & \cellcolor{mygreen}69.76$\pm$6.58 & 1\\
        \midrule
        SISH & 72.39$\pm$10.84 & \cellcolor{mygreen} 70.53$\pm$ 27.91& 57.71$\pm$ 25.90& N/A & 66.88$\pm$7.99 & 3\\
        \midrule
        RetCCL & 34.29$\pm$30.57 & 44.85$\pm$36.92 & 35.60$\pm$ 1.76 & N/A & 38.25$\pm$5.76 & 8\\
        \midrule 
        Yottixel-K\tnote{1} & 64.92$\pm$28.89 & 69.67$\pm$33.27 & \cellcolor{myblue}60.46$\pm$18.02 & N/A & 65.12$\pm$4.76 & 4\\
        \midrule 
        Yottixel-KR\tnote{2} & \cellcolor{myblue}74.53$\pm$17.91 & \cellcolor{myblue}70.00$\pm$31.02 & 59.32$\pm$1.97 & N/A & \cellcolor{myblue}67.95$\pm$7.81 & 2 \\
        \midrule 
        SISH-N\tnote{3} & 58.47$\pm$31.48 & 64.18$\pm$ 32.62& 53.55$\pm$ 15.82& N/A & 58.73$\pm$5.32 & 5\\
        \midrule
        RetCCL-N\tnote{4} & 53.42$\pm$10.44 & 37.55$\pm$32.52 & 36.25$\pm$3.88 & N/A & 42.41$\pm$9.56 & 7\\
        \bottomrule
    \end{tabular}
        \begin{tablenotes}\footnotesize
            \item[1] DenseNet replaced by KimiaNet
            \item[2] DenseNet replaced by KimiaNet, and using ranking after search
            \item[3] SISH with no ranking after search
            \item[4] RetCCL with no ranking after search
        \end{tablenotes}
    \end{threeparttable}
\end{table*}

\paragraph{Key Findings of Search Results Accuracy Measurements (Table \ref{tab:accuracy})}

\begin{quote}
$\square$ \textbf{BoVW} results consistently put it on the 6th rank among the 8 configurations. Considering our modest efforts to train an autoencoder for visual words (to replace handcrafted features), it was surprising to see that BoVW surpasses RetCCL, a recently published method. More surprisingly, BoVW (with 37\%) performs twice better for breasts than SISH (17\%). 

$\square$ \textbf{Yottixel} and its variations consistently achieve the first or second rank for average $F_1$ scores. Only for the MV3 and MV5 of the Skin dataset, Yottixel is less accurate than SISH with statistically insignificant difference (e.g., 70.10\% versus 70.81\%). Yottixel configurations perform best or second best in 8/10 experiments. Yottixel original configuration was the best overall for MV5 retrievals ($\bar{F}_1=69.76\%$) and second best overall for MV3 retrievals ($\bar{F}_1=67.54\%$).

$\square$ \textbf{SISH} delivers the highest score for Skin for MV3 and MV5 (en par with Yottixel-KR) but is consistently ranked in the 3rd or 4th place. That SISH with no ranking algorithm performs poorly has been reported by the SISH authors but a drop of 10\% and more clearly shows that extending Yottixel indexing with an autoencoder does not justify the emphasis on `self-supervised'. SISH authors report improvements over Yottixel up to 45\%. Obviously, we could not reproduce such results for our internal datasets.

$\square$ \textbf{RetCCL} was, with and without the ranking algorithm, the worst search method in these experiments. Strikingly, and in contrast to SISH, RetCCL-N (with no post-search ranking algorithm) performed better than RetCCL. For our challenging breast dataset with 16 rare subtypes, RetCCL completely collapsed (accuracy values of 1 or 2\%). 
\end{quote}

Generally, the accuracy values delivered by all search methodologies are rather low. The highest $F_1$ score is $77.33\%$ for the Liver (ASH vs. NASH vs. normal) delivered by Yottixel for MV5. For Skin the highest values are around $70\%$, and for CRC as low as 65\%. Breast is the most challenging with the highest $F_1$ score of $55.9\%$ by Yottixel-K. Clearly, the accuracy of image search is still below acceptable for clinical utility.

\subsubsection{Indexing \& Search Speed Experiments}

Beyond accuracy, the computational expense to index and search WSIs is a key factor in selecting a capable search technology. 
Table \ref{tab:speed} reports the time efficiency of the four selected search engines. It highlights two crucial aspects: the time taken to create the index for different anatomical sites ($T_{idx}$ in minutes), average indexing time per WSI ($\overline{t}_{idx}$ in minutes) and the average search times ($\overline{t}_{search}$ in minutes) for indexing and conducting the search for a query. The robustness of the search approach (in terms of number of failures) is an additional factor that we measure: being fast is not enough. The Table \ref{tab:speed} also reports how many times each search engine failed to process a WSI. \textcolor{black}{The predominant reason for failures appears to be absence of any patch for processing after pre- or post-processing of patches.}

\begin{table*}[hbt]
    \centering
    \caption{Index creating times $T_{idx}$ (in minutes), average indexing time per WSI $\overline{t}_{idx}$ (in minutes), average search times $\overline{t}_{search}$ (in minutes), and number of fails ($N_\textrm{F}$) for the internal datasets ($R_{idx}$ ranking based on indexing time, $R_{search}$) ranking based on search time, and $R_{F}$ ranking based on failures). The best results are highlighted in green, the second best in blue colors.}
    \label{tab:speed}
    \begin{threeparttable}
    \begin{tabular}{lccccccc}
        \toprule
         &$T_{idx}$ & $\overline{t}_{idx}$ & $R_{idx}$ & $\overline{t}_{search}$ & $R_{search}$ & $N_\textrm{F}$ & $R_{F}$ \\
         & \small [Liver, Skin, CRC, Breast] & & \\
        \midrule
        BoVW & [43, 32 , 43 , 10] & \cellcolor{mygreen} 0.10 & 1 & \cellcolor{myblue} 0.49 & 2 & \cellcolor{myblue}[2,1,0,1]& 2 \\
        \midrule
 
        Yottixel & [27, 95, 57, 15] & \cellcolor{myblue} 0.15 & 2 & \cellcolor{mygreen}0.18 & 1 & \cellcolor{mygreen}[2,0,0,1] & 1\\
        \midrule
        SISH & [152, 984, 706, 257] & 1.65 &  3 &  2.51 & 3 & [166,8,6,2] \tnote{$\mathbf{\dag}$} & 4\\
                \midrule
        RetCCL & [297, 1148, 943, 155] & 2.00 & 4 & 2.53 & 4 & [14,10,6,7] & 3\\
        \midrule 
                
        Yottixel-K\tnote{1} & [27, 95, 57, 15] & \cellcolor{myblue} 0.15 & 2 & \cellcolor{mygreen}0.18 & 1 & \cellcolor{myblue}[2,1,0,1] &  2\\
        \midrule
        Yottixel-KR\tnote{2} & [27, 95, 57, 15] & \cellcolor{myblue} 0.15 & 2 & \cellcolor{mygreen}0.18 & 1 &  \cellcolor{myblue}[2,1,0,1] & 2\\

        \midrule

        SISH-N\tnote{3} & [152, 984, 706, 257] & 1.65 & 3 &  2.51 & 3 &  [166,8,6,2] \tnote{$\mathbf{\dag}$} & 4\\
        \midrule
        RetCCL-N\tnote{4} & [297, 1148, 943, 155] & 2.00 & 4 & 2.53 & 4 & [14,10,6,7] & 3\\
        \bottomrule
    \end{tabular}
    \begin{tablenotes}\footnotesize
        \item[$\dag$] SISH failed to process a total of 14\% of cases
        \item[1] DenseNet replaced by KimiaNet
        \item[2] DenseNet replaced by KimiaNet, and using ranking after search
        \item[3] SISH with no ranking after search
        \item[4] RetCCL with no ranking after search
    \end{tablenotes}
    \end{threeparttable}
\end{table*}

\paragraph{Key Findings of Indexing and Search Speed Measurements (Table \ref{tab:speed})}

\begin{quote}
$\square$ \textbf{BoVW} offers the fastest indexing scheme with a combined 128 minutes for all datasets. For search time (index and search), however, BoVW falls expectedly behind Yottixel due to binary comparisons of Yottixel. BoVW is also quite reliable with only a few failed cases. 

$\square$ \textbf{Yottixel} delivers efficient indexing (ranked 2nd) and fast search (ranked 1st) mainly due to the barcodes it uses. With very few failures, Yottixel is also the most reliable search solution.

$\square$ \textbf{SISH} takes 16 times more time than BoVW to index images which is certainly emanating from autoencoder and codebook generation on top of DenseNet and barcode like Yottixel. SISH is 14 times slower than Yottixel for search. As well, SISH is the most unreliable approach failing in 14\% of cases. 

$\square$ \textbf{RetCCL} is the slowest in indexing. Replacing stain histogram with deep features in Yottixel's mosaic makes RetCCL the most inefficient search engine in indexing. It also fails to process 3\% of the cases.
\end{quote}

Generally, the one-histogram approach and barcoding framework explain the superiority of BoVW and Yottixel. Using any logarithmic or even theoretical constant-time tree structures like SISH will not accelerate the search when indexing is sluggish and violates the basic efficiency rules. Furthermore, the average indexing time per WSI is independent of the archive size. 

\subsubsection{Storage Requirements}

Digital pathology no longer relies on the examination of glass slides; however, it necessitates high-performance digital storage to effectively archive gigapixel WSIs\cite{hanna2022integrating}. A significant portion of the expenses associated with transitioning to a digital system is allocated to the acquisition of high-performance storage solutions  \cite{eccher2023cost}.  This will be a major financial factor for any hospital. The selection of the best search engine without knowing its storage requirement to index WSI archives is practically impossible. 
Table \ref{tab:storage} reports the storage requirements in kilobytes (KB) for indexing and saving WSIs for different datasets. Assuming a linear relationship for all methods, the table also includes an estimate of the storage required for one million WSIs. It's important to consider that the stored index needs to be loaded into memory for the search \& matching to take place.


\begin{table*}[htb]
    \centering
    \caption{Storage required to index and save each WSI [in kilobytes] for the internal datasets ($\bar{S}$ is the average index size for each search method, $\bar{S}_{10^6}$ is the estimate index size for one million WSIs). The best results (lowest storage requirements)
are highlighted in green, the second best in blue colors.}
    \label{tab:storage}
    \begin{threeparttable}
    \begin{tabular}{lcccc|lc|c|c}
        \toprule
         & Liver & CRC & Skin & Breast & index size/WSI & $\bar{S}$ & $\hat{S}_{10^6}$ & $R_{storage}$\\
        \midrule
        BoVW & 8 &15 &15 &2 & \cellcolor{mygreen}0.03 KB & \cellcolor{mygreen} 10 KB &  $\approx$ 10 GB & 1\\
        \midrule 
        Yottixel &  74 & 177 &  102 &  125  & \cellcolor{myblue}0.38 KB & \cellcolor{myblue}120 KB & $\approx$ 119 GB & 2\\
        \midrule
        SISH  & 29320&   23634 &  9045 &   61734 & 97.50 KB & 30933 KB & $\approx$ 31 TB\tnote{$\dag$} & 4 \\
        \midrule
        RetCCL & 1934 &  4378 &  538 &  455 & 5.76 KB & 1826 KB& $\approx$ 1.8 TB & 3\\
        \bottomrule
    \end{tabular}
        \begin{tablenotes}\footnotesize
        \item[$\dag$] We assumed the \emph{best case} scenario for space complexity of the vEB tree in SISH to be implemented and able to reach a linear upper bound.
    \end{tablenotes}
    \end{threeparttable}
\end{table*}



\begin{quote}
    \textbf{Key findings of Storage Requirement (Table \ref{tab:storage})}

    $\square$ BoVW requires the least amount of storage for indexing and saving each WSI which makes BoVW an efficient choice for applications with limited storage resources.
    
    $\square$ Yottixel exhibits a relatively low storage requirement for WSIs. This makes it a storage-efficient solution, particularly for tasks with large datasets.
           
    $\square$ SISH requires the highest storage capacity for indexing WSIs. This high storage demand might pose challenges for applications with limited storage resources.
    
    $\square$ RetCCL also demands substantial storage space. This makes it less storage-efficient, especially for applications dealing with a large number of WSIs.

    $\square$ Both SISH and RetCCL require around 30 to 50 times more storage than Yottixel and BoVW. 
    
\end{quote}

Perhaps the most important measure for the storage (and naturally for the speed) of each search method is the average index size per WSI. With 0.03 kilobytes per WSI, BoVW is the most efficient search approach. SISH needs $\approx$98 kilobytes per WSI which clearly is a tremendous overhead for any digital archive, and impractical for loading into memory to conduct search.

\subsection{Comparative Experiments for External Data}
In this section, we report comprehensive results for search accuracy, indexing and search speed,
and storage requirements of search engines on three public datasets. As BoVW and RetCCL exhibit low accuracy, we restrict the external experiments on Yottixel, SISH, and their variants. 

\subsubsection{Search Accuracy Results}
For further evaluation of the search engines, we apply them on three well-known WSI-level public datasets: CAMELYON16 \cite{DataCAMELYON16} (short CAM16), BRACS \cite{DataBRACS}, and PANDA \cite{DataPANDA}. As mentioned before, due to the extensive nature of leave-one-out validation, we restricted these experiments only using MV@5 and only to SISH and Yottixel as BoVW and RetCCL displayed low accuracy. Table \ref{tab:MV5Public} shows the results. Yottixel-K is the most accurate method for CAM16 and BRACS. It is also the best method with an average $F_1$ score of 56.3\%. SISH lands in the second place with an average $F_1$ score of 52.9\%. However, as Table \ref{tab:speedPublic} shows, SISH fails in 57\% of cases (for 569/1000 WSIs) in the PANDA dataset.

\begin{table*}[htbp!]
    \centering
    \scriptsize
    \caption{Results for MV@5 on the public datasets: CAM16, BRACS, and PANDA.}
    \label{tab:MV5Public}
    
    \begin{threeparttable}
    \begin{tabular}{lccc|ccc|cc}
        \toprule
    &CAM16 \cite{DataCAMELYON16}&BRACS \cite{DataBRACS}& PANDA \cite{DataPANDA}& $\bar{F}_1$ & SE$_R^{MV5}$\\
    \midrule
    Yottixel&64.5$\pm$19.1	&\cellcolor{myblue}52.6$\pm$9.0&	31.8$\pm$12.4&49.6$\pm$16.5 & 3\\
        \midrule
        SISH	&		\cellcolor{myblue}73.5$\pm$13.4&	49.0$\pm$8.9	&\cellcolor{mygreen}36.2$\pm$20.2 \tnote{$\ddagger$} & \cellcolor{myblue}52.9$\pm$19.0 \tnote{$\ddagger$} & 2\\

        \midrule 
        Yottixel-K	&	\cellcolor{mygreen}75.5$\pm$13.4&	\cellcolor{mygreen}59.7$\pm$4.7&	33.8$\pm$12.0 &\cellcolor{mygreen}56.3$\pm$21.0 & 1\\

        \midrule 
        Yottixel-KR	&	\cellcolor{myblue}73.5$\pm$13.4	&51.7$\pm$5.0	&\cellcolor{myblue}35.5$\pm$11.0&45.0$\pm$8.50 & 5\\

        \midrule 
        SISH-N	&		59.0$\pm$28.3&	51.0$\pm$5.0	&29.0$\pm$14.6&46.3$\pm$15.5 & 4\\

        \bottomrule
    \end{tabular}
    \begin{tablenotes}\tiny
            \item[$\ddagger$] SISH failed in search for almost 57\% of PANDA images (see Table \ref{tab:speedPublic})
        \end{tablenotes}
    \end{threeparttable}
\end{table*}







\subsubsection{Search Speed Benchmark}
Table \ref{tab:speedPublic} reports the time efficiency of the selected search engines. It highlights two crucial
aspects: the time taken to create the index for different anatomical sites ($T_{idx}$ in minutes) and
the average search times ($T_{search}$ in seconds) for conducting searches. It also demonstrates
how many times each search engine failed to process a WSI among datasets. Note, the search time
encompasses both the indexing time and the matching time. Table \ref{tab:speedPublic} confirms what we reported for our internal (private) dataset in Table \ref{tab:speed}. SISH is much slower and fails more often than Yottixel.

\begin{table*}[htb]
    \centering
    \caption{Index creating times $T_{idx}$ (in minutes), average indexing time per WSI $\overline{t}_{idx}$(in minutes), average search times $\overline{t}_{search}$ (in minutes), and number of fails ($N_\textrm{F}$) for public dataset.}
    \label{tab:speedPublic}
    
    \begin{tabular}{lccccccc}
        \toprule
         &$T_{idx}$ & $\overline{t}_{idx}$ & $R_{idx}$ & $\overline{t}_{search}$ & $R_{search}$ & $N_\textrm{F}$ & $R_{F}$ \\
         & \tiny [CAM16, BRACS, PANDA] & 1.22 & & \\
        \midrule

        Yottixel & [27,22,14](63) & \cellcolor{mygreen} 0.05 &  1 & \cellcolor{mygreen} 0.16 & 1 & \cellcolor{mygreen}[1,2,14] & 1\\
        \midrule
        SISH & [448,254,77](779) & 1.22 & 2 &  2.36 & 2 & \cellcolor{myblue}[0,1,569] & 2\\
                \midrule

        Yottixel-KR\tnote{2} & [28,24,14](66) & \cellcolor{mygreen} 0.05 & 1 & \cellcolor{mygreen} 0.16 & 1 &  \cellcolor{mygreen}[1,2,14] & 1\\

        \midrule

        SISH-N\tnote{3} & [448,254,77](779) & 1.22 & 2 &  2.36 & 2 &  \cellcolor{myblue}[0,1,569] & 2\\
        \bottomrule
    \end{tabular}
 
\end{table*}

\subsubsection{Storage Benchmark}
Table \ref{tab:storagePublic} presents storage requirements in kilobytes (KB) for indexing and saving WSIs for different datasets. It lists the storage requirements for various methods across all datasets. The table
also includes an estimate of the storage required for one million WSIs. The results for the three public datasets are aligned with the results we reported for our internal (private) datasets in Table \ref{tab:storagePublic}. The indexing overhead of SISH is prohibitively large (165 kilobytes per WSI) whereas Yottixel is rather lean (0.23 kilobytes per WSI).

\begin{table*}[htb]
    \centering
    \caption{Storage required to index and save each WSI [in kilobytes] for the external datasets ($\bar{S}$ is the average index size per dataset, $\bar{S}_{10^6}$ is the estimate index size for one million WSIs). The best results (lowest storage requirements)
are highlighted in green, the second best in blue colors.}
    \label{tab:storagePublic}
    \footnotesize 
    \begin{threeparttable}
    \begin{tabular}{lccc|cc|cc}
        \toprule
         & CAM16 \cite{DataCAMELYON16}&BRACS \cite{DataBRACS}& PANDA \cite{DataPANDA} & $S$/WSI & $\bar{S}$ & $\bar{S}_{10^6}$ & $R_{storage}$ \\
        \midrule 
        Yottixel &  179 KB &  169 KB &  67 KB  & 0.23 KB & 138 KB & $\approx$ 138 GB & 1\\
        \midrule
        SISH  & 39997 KB &    55124 KB &   12109 KB & 165.73 KB& 35743 KB & $\approx$ 36 TB & 2 \\
        \bottomrule
    \end{tabular}
    \end{threeparttable}
\end{table*}

\subsection{Overall Performance Ranking for Internal \& External Data}

To compare and rank the performance of the four selected search engines in a holistic manner, considering multiple aspects of their performance, results from all benchmarks were used. The reviewed performance metrics include scores for different aspects of the search, including top-1 accuracy ($R_{top1}$),  average $F_1$ score at 3 ($R_{MV3}$),  average $F_1$ score at 5 ($R_{MV5}$), indexing time ($R_{idx}$), search time ($R_{search}$), the number of failed WSIs ($R_{F}$), and the storage requirement ($R_{storage}$).

Table \ref{tab:ranksinternal} presents an average performance rank for internal datasets, denoted as $R$ based on these individual SE$_R$ values. Each row corresponds to a specific search engine or method, such as Yottixel-KR, Yottixel, Yottixel-K, BoVW, SISH-N, SISH, RetCCL-N, and RetCCL. The table provides rankings for each search method across the various $_R$ metrics. The $R$ values (see equation 1) provide a composite ranking, allowing readers to identify which method performs better overall, taking into account a range of performance measures. 

Yottixel's original configuration and its other two variants secure the top three positions when compared to other tested search engines, underscoring its accuracy, effectiveness, and robustness. Although BoVW demonstrated poor performance in search accuracy, its time and storage efficiency improved its overall ranking to 3.57 to earn the 4th place. SISH, with and without ranking, shows mediocre performance in all benchmarks placing it after Yottixel and BoVW. The number of failures and inefficient memory requirements also play a role in earning a rank of 3.86 among 8 scenarios. RetCCL, with and without ranking, ranks the lowest among the tested search engines in all areas.

\begin{table*}[htb]
    \centering
    \caption{Overall \textbf{Performance for Internal Datasets}: Based on all previous ratings (for top-1, MV3, MV5, indexing time, search time, number of failed WSIs, and required storage), we calculate $R^*_\textrm{internal}$ as the average performance rank for each search method.   }
    \label{tab:ranksinternal}
    \begin{threeparttable}
    \begin{tabular}{lccccccc|c}
          & $R_{top1}$ & $R_{MV3}$ & $R_{MV5}$ & $R_{idx}$ & $R_{search}$ & $R_{F}$ & $R_{storage}$ & $R^{*}_{\textrm{internal}}$\\
          \midrule
Yottixel \tnote{$\ddagger$}   & 3	& 2	& 1	& 2	& 1	& 1	& 2	& \textbf{1.71}\\ \midrule
Yottixel-KR \tnote{2} & 2	& 1	& 2	& 2	& 1	& 2	& 2	& \textbf{1.71}\\ \midrule
Yottixel-K \tnote{1}  & 1 & 3	& 4	& 2	& 1 & 2	& 2	& \textbf{2.14}\\ \midrule
BoVW        & 6	& 6	& 6	& 1	& 2	& 2	& 1	& \textbf{3.43}\\ \midrule
SISH \tnote{$\ddagger$}       & 4	& 4	& 3	& 3	& 3	& 4	& 4	& \textbf{3.57}\\ \midrule
SISH-N \tnote{4}      & 5	& 5	& 5	& 3	& 3	& 4	& 4	& \textbf{4.14}\\ \midrule
RetCCL-N \tnote{5}   & 8	& 8	& 7	& 4	& 4	& 3	& 3	& \textbf{5.28}\\ \midrule
RetCCL \tnote{$\ddagger$}     & 7	& 7	& 8	& 4	& 4	& 3	& 3	& \textbf{5.43}\\
        \bottomrule
        \end{tabular}
        \begin{tablenotes}\footnotesize
            \item[$\ddagger$] As originally proposed  
            \item[1] DenseNet replaced by KimiaNet
            \item[2] DenseNet replaced by KimiaNet, and using ranking after search
            \item[3] SISH with no ranking after search
            \item[4] RetCCL with no ranking after search
            \item SE$_R$ = search engine ranking in each benchmark
        \end{tablenotes}
    \end{threeparttable}
\end{table*}

Table \ref{tab:ranksexternal} shows the same ratings for the external dataset. As the two more accurate methods (compared to BoVW and RetCCL), Yottixel and SISH were only tested for MV5. The results confirm the general rating that Yottixel is more accurate, more efficient, and more reliable than SISH. The only difference with internal results is that Yottixel-K climbs to the first place ahead of Yottixel. This means that even for a method that is generally reliable, the selection of a deep network may still need some trial-and-error for specific datasets.

\begin{table*}[htb]
    \centering
    \caption{Overall Performance for External Datasets: Based on all previous SE$_R$ (for top1, MV@3, MV@5, indexing time, search time, number of failed WSIs, and required storage), we calculate $R$ as the average performance rank for each search method.   }
    \label{tab:ranksexternal}
    \begin{threeparttable}
    \begin{tabular}{lccccccc|c}
          & $R_{MV5}$ & $R_{idx}$ & $R_{search}$ & $R_{F}$ & $R_{storage}$ & $R^*_{external}$\\
        \midrule
                Yottixel-K\tnote{1} & 1 & 1 & 1 & 1 &  1 & \textbf{1.00} \\
        \midrule
                \textbf{Yottixel}\tnote{$^\ddagger$} & 3 & 1 & 1 & 1  & 1 & \textbf{1.40} \\
                \midrule 

        Yottixel-KR\tnote{2} & 5 & 1 & 1 & 1 & 1 & \textbf{1.80} \\
        \midrule
        \textbf{SISH}$^\ddagger$  & 2  & 2 & 2 & 2 & 2 & \textbf{2.00} \\
        
        \midrule
        SISH-N\tnote{3}  & 4 & 2 & 2 & 2 & 2 & \textbf{2.60}\\
        \bottomrule
               \end{tabular}
        \begin{tablenotes}\footnotesize
            \item[$\ddagger$] As originally proposed  
            \item[1] DenseNet replaced by KimiaNet
            \item[2] DenseNet replaced by KimiaNet, and using ranking after search
            \item[3] SISH with no ranking after search
            \item[4] RetCCL with no ranking after search
        
        \end{tablenotes}
    \end{threeparttable}
\end{table*}

\section{Discussions}
Benchmarking image search engines in histopathology is a critical process that facilitates the identification of both strengths and weaknesses of available technologies. By evaluating the performance of different image search engines, researchers and practicing pathologists gain valuable insights into the capabilities and limitations of these technologies. A rigorous analysis not only enhances our understanding of their operational efficiency but also guides the refinement and innovation of new approaches for tissue image search.

The \textbf{top-1 search accuracy} is a metric often used to assess the correctness of the top-matched image in search engines. However, it is important to recognize that the top-1 search accuracy treats the search process as a form of classification, which can be a somewhat limited measure, particularly because these search engines typically lack a training phase as seen in traditional classification models. Nonetheless, when we are dealing with methods that can perform direct comparisons between WSIs, achieving a high top-1 match rate can be a robust indicator of how effectively a WSI representation is generated. This highlights the importance of not just classifying images accurately but also ensuring that the generated representation is suitable for making these one-to-one comparisons. 

\textcolor{black}{The impact of deep features on search accuracy is expected to be high. We know that stain variations \cite{babaie2019deep,otalora2019staining,de2021deep} and data bias \cite{dehkharghanian2023biased,hagele2020resolving} may affect the equality of embeddings. One may design and train a general model feature extraction in histopathology. \textcolor{black}{For instance, Virchow model used one million WSIs for its training \cite{vorontsov2023virchow}.} As well, one might think in direction of customized training customized networks for specific organs and diseases \cite{ba2014deep,tizhoosh2024image}. }

Furthermore, evaluating \textbf{top-3 and top-5 search accuracy} (by taking the majority votes MV3 and MV5) provides insight into the consistency of retrieved results, as they essentially reflect the most frequently matched WSIs. Interestingly, across all four tested search engines, the results between top-1, MV3, and MV5 search accuracy remained almost consistent. Notably, Yottixel, particularly in its original configuration and two other variants, often secured the top three positions when compared to other tested search engines. BoVW, while showing impressive time and storage efficiency, demonstrated a relatively poor performance in search accuracy. Considering our modest effort in training an autoencoder to learn visual words this was to be expected. SISH, whether with or without ranking, displayed mediocre performance in search accuracy, placing it behind Yottixel. Finally, RetCCL, both with and without ranking, consistently delivered the lowest accuracy among the tested search engines.

While achieving top-tier search accuracy is undeniably vital, it is equally essential for these systems to demonstrate \textbf{efficiency in terms of time and storage} utilization for practical implementation. Notably, Yottixel and BoVW have proven to be efficient in storage and execution time.  Conversely, SISH and RetCCL have lagged behind, showing sluggish indexing behaviour and impractical storage needs, which considerably limits their scalability and broader adoption. Particularly, SISH has excessive storage need due to using vEB trees, a factor that makes SISH slow for small and medium size datasets, and infeasible for large datasets. While RetCCL can readily enhance its performance by substituting deep features for mosaic building, SISH faces a significant challenge due to prolonged indexing times required to supply its tree structure with long integers.

Another pivotal takeaway from our validations is that an extra \textbf{ranking algorithm outside the search engine} is quite problematic. It seems applying of ranking to the search results has been motivated by low-quality results. The removal of ranking from RetCCL and SISH had intriguing consequences, revealing nuances in their performance. This alteration led to a deterioration in the search capabilities of SISH, confirming that ranking played a crucial role in refining their search accuracy. As for RetCCL, the ranking delivers more inconsistent results. On the contrary, Yottixel only experienced a rather slight improvement with the addition of ranking, hinting at its already consistent and reliable results with no further post-processing (we only added ranking after Yottixel for the sake of fair comparison and completeness of experiments, a step that has been omitted in SISH paper). 

As previously discussed, the critical task of \textcolor{black}{\textbf{patient tissue matching}} through matching WSI to other WSIs underscores the need for an efficient generation of WSI representations (like Yottixel's mosaic) and a reliable scheme for WSI-to-WSI comparison (like Yottixel's median-of-minimum distance calculation). This is where SISH and RetCCL fall short, as they not only do not possess the capability to generate expressive WSI representations, but also they lose the capability of WSI-to-WSI matching. The former shortcoming for SISH is due to crippling Yottixel's barcodes with a patch-oriented tree structure and a post-search ranking scheme and for RetCCL due to the adoption of the SISH ranking scheme. The latter shortcoming is mainly due to using a redundant ranking algorithm after the search. \textcolor{black}{Both the vEB tree and the post-search ranking separate mosaic patches, rendering mosaic-to-mosaic comparison infeasible.}


A major observation through our investigations was the \textbf{lack of novel patching} algorithms. BoVW has a unique divide strategy that enables WSI representations through the histogram of visual words. Yottixel is the only recent search engine that introduced a novel framework for patching, coined ``mosaic'', a two-stage stain/proximity clustering followed by intra-cluster sampling. SISH and RetCCL use Yottixel's mosaic for patching. There is an urgent need for novel ideas for WSI patching to improve search results and the efficiency of indexing.  

\vspace{0.1in}

Our data collection and engagement using the four methods extended over several months. Beyond the accuracy, speed, storage, and robustness numbers that we have reported, we can conclude

\begin{quote}
$\blacksquare$ \textbf{BoVW} and \textbf{Yottixel} present comprehensive search solutions with innovative concepts. Their combination of high speed and efficient storage is highly valuable. However, to enhance their accuracy, integration with a well-trained backbone network and the implementation of adjustments to the primary site are necessary. As Yottixel is a commercial product, researchers may have more \emph{freedom} to investigate different variants of BoVW.

$\blacksquare$ \textbf{Yottixel}  relies on manual settings for its mosaic, particularly regarding the number of clusters (i.e., tissue types) and the percentage of sampling within clusters. Yottixel, along with SISH and RetCCL using the same scheme, would greatly benefit from a fully automated patch selection. Additionally, implementing a logarithmic approach to barcode comparison could further enhance the speed of median-of-minimum Hamming distance calculations, making Yottixel even faster than it already is.

$\blacksquare$ \textbf
{SISH}, a Yottixel variant, deviates from the principles of Occam's Razor, introducing speed and scalability issues. The attempt to enhance Yottixel has led to unnecessary complexity in SISH, marked by trial and error. A fundamental design flaw is evident as SISH relies on vEB trees, assuming $\mathcal{O}(\lg \lg N)$ time, but neglecting the exponential space requirements $\mathcal{O}(2^m)$. The use of long integers in this outdated structure, coupled with massive overhead, renders SISH impractical for large $N$, making the loading and processing of terabytes of data into memory infeasible. Regardless of archive size, SISH exhibits sluggish indexing due to the necessity of generating very long integers for the tree.

$\blacksquare$ \textbf{RetCCL} barely qualifies as a search engine. \textcolor{black}{Its primary focus is on the CCL network, which appears to provide non-expressive embeddings for tissue morphology when used for image retrieval}.

$\blacksquare$ \textcolor{black}{\textbf{All four methods} can incrementally index new WSIs although they may then perform poor if the new tissue pattern does not fall inside the latent space of the deep network used.}
\end{quote}

These findings emphasize the trade-offs and varied strengths and weaknesses among different image search approaches in histopathology, emphasizing the complex interplay between search accuracy, speed, and memory efficiency in this field. Beyond these requirements, we have also observed the following deficiencies:  

\begin{itemize} 
    \item The accuracy level of all search engines is still low for clinical utility.

    \item  We should prioritize both speed and storage considerations in digital pathology.
    
    \item No automated curation framework exists for WSI selection.
    
    \item No algorithm has been put forward for setting magnification level and patch size.
    
    \item  No multimodal search approach has been proposed.
\end{itemize}


\section*{Data Availability}
The internal datasets analyzed during the current study are not publicly available due to privacy concerns and ethical considerations. 

\section*{Code Availability}
\textcolor{black}{The Yottixel’s code, and KimiaNet model and its weights are available at Kimia Lab’s GitHub (https://github.com/KimiaLabMayo). SISH code is available at a GitHub repository (https://github.com/mahmoodlab/SISH). RetCCL code and model are available at a GitHub repository (https://github.com/Xiyue-Wang/RetCCL). BoVW code is available at Kimia Lab’s GitHub (https://github.com/KimiaLabMayo).}

\section*{Disclosures}
\textcolor{black}{Hamid R. Tizhoosh is the inventor of Yottixel’s barcoding, a patented technology that is owned by Huron Digital Pathology (St. Jacobs, ON, Canada) based on sponsorship agreements. Hamid R. Tizhoosh has no relationship with Huron Digital Pathology since December 2020.}

\bibliographystyle{unsrt}
\bibliography{bibliography}


\end{document}